\documentclass[aps,prl,reprint,groupedaddress]{revtex4-2}
\usepackage{graphicx}
\usepackage{color}
\usepackage[draft]{hyperref}
\usepackage{amsmath}
\newcommand{\dcor}[1]{\color{black}{#1}}
\usepackage[normalem]{ulem}

\begin{document}

% Use the \preprint command to place your local institutional report 
% number in the upper righthand corner of the title page in preprint mode.
% Multiple \preprint commands are allowed.
% Use the 'preprintnumbers' class option to override journal defaults
% to display numbers if necessary
%\preprint{}

%Title of paper
\title{
%Observation and characterization 
Observation of stochastic resonance in directed propagation of cold atoms}

% repeat the \author .. \affiliation  etc. as needed
% \email, \thanks, \homepage, \altaffiliation all apply to the current
% author. Explanatory text should go in the []'s, actual e-mail
% address or url should go in the {}'s for \email and \homepage.
% Please use the appropriate macro foreach each type of information

% \affiliation command applies to all authors since the last
% \affiliation command. The \affiliation command should follow the
% other information
% \affiliation can be followed by \email, \homepage, \thanks as well.
\author{Alexander Staron$^{1, \dag}$, Kefeng Jiang$^{1, \dag}$, Casey Scoggins$^1$, Daniel Wingert$^{1}$, David Cubero$^{2}$, and Samir Bali$^{1}$}
\email[Corresponding author: ]{balis@miamioh.edu} 
%\homepage[]{Your web page}
%\thanks{}
%\altaffiliation{}
\affiliation{$^{1}$ Department of Physics, Miami University, Oxford, Ohio 45056-1866,
USA \\ $^{2}$ Departamento de Fisica Aplicada I, Universidad de Sevilla, Spain 
$^{\dag}$These two authors contributed equally
}

%Collaboration name if desired (requires use of superscriptaddress
%option in \documentclass). \noaffiliation is required (may also be
%used with the \author command).
%\collaboration can be followed by \email, \homepage, \thanks as well.
%\collaboration{}
%\noaffiliation

\date{\today}

\begin{abstract}
\textcolor{black}{
Randomly diffusing atoms confined in a dissipative optical lattice are illuminated by a weak probe of light.
%, of intensity less than 1\% of the total lattice intensity.
%A weak probe of light illuminates randomly diffusing atoms confined in a dissipative optical lattice.
%We illuminate randomly diffusing atoms confined in a dissipative optical lattice with a weak probe beam. 
The probe transmission spectrum reveals directed atomic propagation that occurs perpendicular to the direction of probe beam propagation.
%of a specific velocity-class of atoms. 
Resonant enhancement of this directed propagation is observed as we vary the random photon scattering rate. We experimentally characterize this stochastic resonance as a function of probe intensity and lattice well depth. A simple model reveals how the probe-excited atomic density waves and optical pumping rates conspire to create directed atomic propagation within a randomly diffusing sample.}
%By illuminating a dissipative optical lattice with a weak frequency-scanning probe beam and detecting the probe transmission spectrum we observe a resonant enhancement in the directed propagation of cold atoms as we vary the rate of random photon scattering. The directed atomic propagation is induced by probe intensities less than 1\% of the total lattice intensity, and occurs perpendicular to the direction of probe beam propagation. 
%%Photon scattering disrupts Hamiltonian atomic motion in the reactive potentials and is thus viewed as noise, analogous to Brownian fluctuations in a thermal system.
%%The resonant response of the system as a function of random 
%%environmental noise 
%%photon scattering is a signature of stochastic resonance.  
%We experimentally characterize this stochastic resonance as a function of the probe intensity and the lattice well-depth. A simple one-dimensional model reveals how the 
%{\dcor probe-excited atomic density waves}
%%\sout{probe-modified ground state potentials} 
%and optical pumping rates
%%light-shifted ground state lattice potentials and the optical pumping rates, both modified by the probe, 
%%probe-induced modifications of the light-shifted ground state lattice potentials on one hand and the optical pumping rates on the other 
%conspire to create directed atomic propagation within a randomly diffusing sample. 

%(531 characters; reduce $<$ 600)
%, 617 char w spaces, 87 words)
\end{abstract}
%REDUCE TO 600 CHARACTERS

% insert suggested keywords - APS authors don't need to do this

%\keywords{}

%\maketitle must follow title, authors, abstract, and keywords
\maketitle
\textcolor{black}{Random fluctuations dominate the transport of sub-microscopic systems immersed in a noisy environment, e.g., 
spontaneous emission recoils in the case of resonantly illuminated cold atoms. }
%in the case of biomolecular motors by thermal collisions with surrounding water molecules, or 
The ability of ``Brownian ratchets" to convert random fluctuations into useful directed motion is a central topic in non-equilibrium statistical physics that has been carefully explored in theory and experiment~\cite{cuberobook,hanggi1,reimann}. In particular, the phenomenon of ``stochastic resonance", which refers to a peak in system response as a function of increasing noise strength, has received wide attention in the physics community~\cite{wellens,gamaitoni,wiesenfeld},
%First introduced in 1981 to explain the periodic recurrence of ice ages~\cite{benzi,nicolis}, stochastic resonances have since been observed in diverse systems in physics, biology~\cite{plos} and engineering~\cite{scirep}, ranging from bistable ring-laser cavities~\cite{rroy} to neural networks~\cite{moss,jung} and human transcranial processes~\cite{2011brain}, and have even been applied toward preventing falls in the elderly~\cite{2019falls} and socio-economic modeling~\cite{socio}.
even finding application in climate science~\cite{climate}, biology~\cite{plos} and, more recently, in engineering~\cite{scirep,natcomm}. 
It has been pointed out that naturally occurring protein motors are able to power the processes of life by harnessing energy from surrounding Brownian fluctuations with efficiencies that are orders of magnitude larger than any artificial nano machine built to date~\cite{lewandowski}. 
\textcolor{black}{Thus the notion that the controlled addition of random noise fluctuations may help rather than hinder system performance has important implications for optimizing the efficiency of nano devices~\cite{pekola2015} and sensors~\cite{nat2019,nat2021} 
in situations where environmental noise is significant.}

Cold atoms confined in {dissipative} optical potentials~\cite{grynreview}, where spontaneous emission is significant, 
%formed by intersecting laser beams tuned near atomic resonance 
are an ideal testbed to study stochastic resonance:
%Here, the system is the confining potential and the random environmental noise could be spontaneous emission or thermal collisions - the system and environment 
The
%in contrast to solid-state or biomimetic architectures, 
system (i.e., the confining potential) and the random environmental noise (i.e., spontaneous emission recoils, or thermal collisions) 
can be {independently} 
%and precisely and widely 
varied 
%over a wide range 
by adjusting beam parameters. 
Recently, 
in an elegant experiment, 
stochastic resonances were detected in cold atoms in a dissipative double-well potential formed by splitting a magneto-optical trap with a blue-detuned sheet of light: Random thermal collisions caused the atoms to hop between the two wells, and 
%the authors observed
%were able to observe thermal activation based 
stochastic resonances were observed in the interwell hopping rate as a function of temperature, while also varying barrier height and atom number~\cite{hume}. 
%, and a theoretical prediction~\cite{grynbergprl} followed by detection~\cite{grynbergepl} of stochastic resonance in the directed propagation of atoms in periodic potentials. 
%caused by asymmetric modulation of the lattice wells

In this work, we report on the observation and experimental characterization of 
%\textcolor{black}{spontaneous emission-mediated} 
stochastic resonance in the directed propagation of cold atoms confined in a \textcolor{black}{dissipative} optical lattice. Here, transfer 
%hopping 
between adjacent wells of the periodic potential array is caused by \textcolor{black}{stochastic optical pumping processes}. 
In previous work, stochastic resonance in a dissipative lattice \textcolor{black}{was predicted~\cite{grynprl2002,grynpra2003},} and preliminary evidence was observed by
%Preliminary evidence of stochastic resonance in a dissipative lattice has been observed previously by 
%asymmetrically 
modulating the lattice potential 
%to instigate directed propagation~\cite{grynprl1996,grynprl2002}, 
and detecting a resonant enhancement in small center-of-mass displacements of the diffusing atom cloud~\cite{grynbergepl}. 
In contrast, we reveal stochastic resonance via pump-probe spectroscopy, 
%which is far more rapid, robust, and significantly less data intensive. 
%where the lattice itself serves as the pump. Compared to diffusion measurement pump-probe spectroscopy is rapid and robust permitting us to 
which permits a first \textcolor{black}{experimental exploration of the dependence of the stochastic resonance on the lattice well-depth and the strength of lattice modulation.} {\dcor A new theory \cite{cubero22}, based on the decomposition of the current into its atomic density wave contributions},
%\sout{A new physical picture, using a $F_{g} = 1/2 \leftrightarrow F_{e} = 3/2$  atom randomly diffusing} 
elucidates how probe-modulated ground-state potentials and optical pumping rates conspire to generate resonantly enhanced directed propagation.
%Reinhardt: The sentence "A new physical picture is developed, using a 1/2 -> 3/2 atom randomly diffusing..." is a bit confusing.  You mean an atom being driven between states with F=1/2 and F=3/2? A: I now specify $F_{g} = 1/2 \leftrightarrow F_{e} = 3/2$...is this enough?
%\textcolor{black}{POLISH!}
%EXPLAIN EXACTLY / CLEARLY 1) DIRECTED VS DIFFUSION, THEN 2) RESONANTLY ENHANCED DIRECTED

{\dcor We consider cold atoms in a so-called  3D-lin$\perp$lin configuration \cite{grynreview,grynpra2003}, formed by confining the  $F_{g} = 1/2 \leftrightarrow F_{e} = 3/2$ atoms in an optical lattice created by the superposition of four red-detuned laser beams {\textcolor{black}{ of amplitude ${\cal E}_0$ and frequency $\omega_L$. Two beams lie}} in the plane {\textcolor{black}{$xOz$}}, with angle $\theta_x$, counterpropagating to the other two, which are in the plane {\textcolor{black}{$yOz$}} with angle $\theta_y$, as illustrated in Fig.~\ref{fig:fig1}~(a). {\textcolor{black}{The beams produce 
two light-shifted ground-state $\pm 1/2$-spin potentials $U_\pm(x,y,z)$ with well-depth $U_0$ shown in Fig.~\ref{fig:fig1}(b)}} (see Appendices A and B).
%%%%%%%%%%%%%%%%%%%
\begin{figure}[b]
\includegraphics[width=8.4cm]{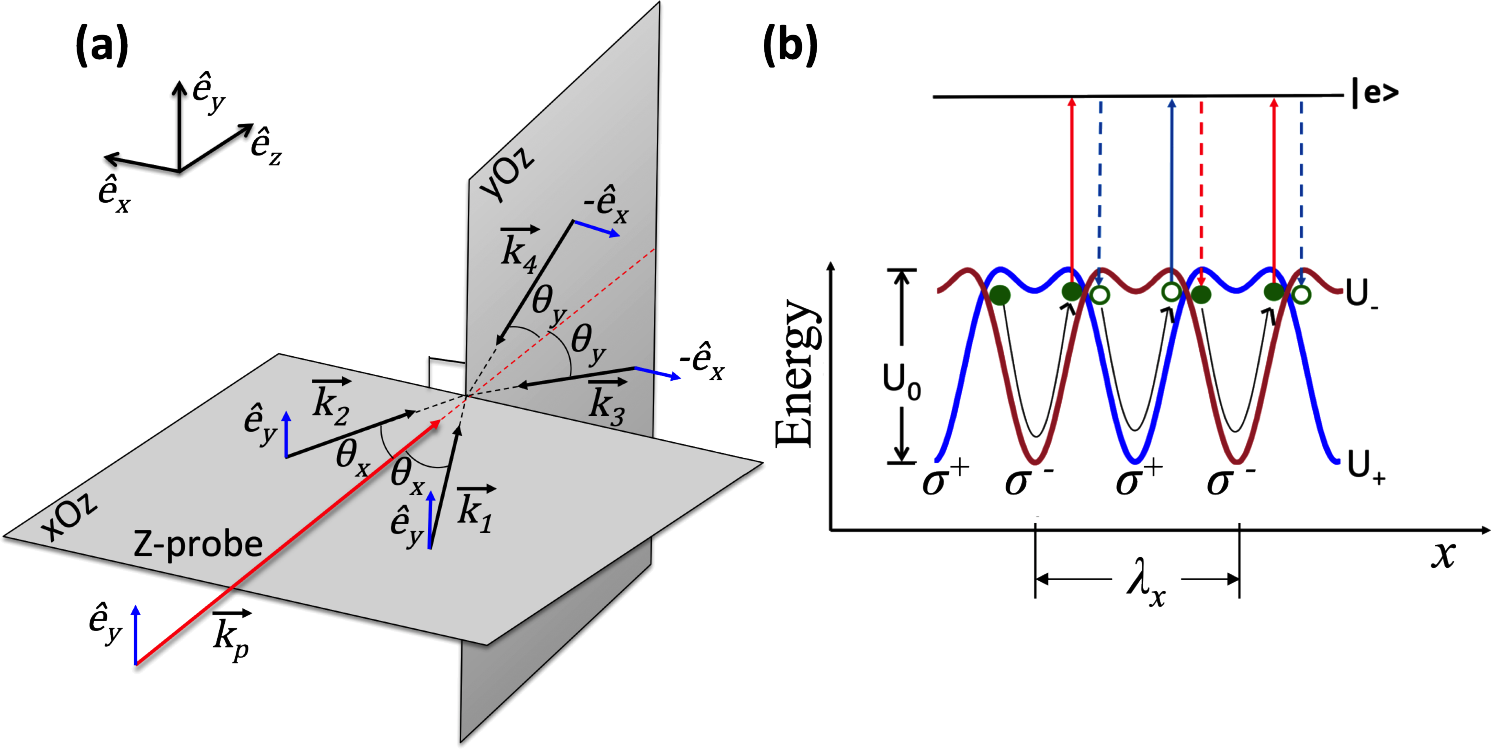}
%\vspace{-3mm}
\caption{\label{fig:fig1}
\dcor
(a)  Scheme of the 3D tetrahedral lin$\perp$lin lattice \newline
(b) Optical potential {\textcolor{black}{$U_\pm$}} along the $x$-direction ($y=z=0$) seen by a $F_{g}=1/2 \rightarrow F_{e}=3/2$ atom in the above lattice, {\textcolor{black}{with well-depth $U_0$ and spatial periodicity $\lambda_x = 2\pi/k_x$}}. 
%[TO DO: please make panel (b) and provide panel (a) {\textcolor{green}{in vectorial version}} (pdf) ]
}
%\vspace{-2mm}
\end{figure}
%%%%%%%%%%%%%%%%%%%
%\begin{equation}
%U_\pm(x,y,z)=U_0\left[-\cos(2 k_x x) - \cos(2 k_y y) \pm 2 \cos(k_x x) \cos(k_y y) \cos( k_z z) \right],
%\end{equation}
%where $k_x=k\sin(\theta_x)$,  $ k_y=k\sin(\theta_y)$, $k_z=k[\cos(\theta_x) + \cos(\theta_y)]$, $U_0=-2\hbar \Delta'_0/3 $. 
We focus on movement along the $x$-direction. A one-dimensional model arises after neglecting movement in the other perpendicular directions \cite{grynprl1996,grynprl2002,grynbergepl,grynpra2003}. By formally considering $y=z=0$, the optical potential in each ground state is given by  \cite{grynreview,grynpra2003}
\begin{equation}
U_\pm(x)=\frac{U_0}{4}\left[-3-\cos(2 k_x x) \pm 2 \cos(k_x x) \right],
%% U0_thispaper = 4*U0_sim
\label{eq:Upm}
\end{equation}
with transition probability rates between them given by
\begin{equation}
\gamma_\pm(x)=\frac{2\Gamma_S}{9}\left[ 3+\cos(2 k_x x)\pm 4\cos(k_x x)\right],
\label{eq:gammapm}
%c gammamp=1/9 Gammap (2+Cos[2 Kx x]+Cos[2 Ky y]-4 Cos[Kx x] Cos[Ky y] Cos[(cosx+cosy) k z])
%c gammapm=2/9 Gammap (Cos[Kx x]^2+Cos[Ky y]^2+2 Cos[Kx x] Cos[Ky y] Cos[(cosx+cosy) k z])
%c     cosx=cos(thetax)   cosy=cos(thetay)  Kx=k*sin(thetax)   ky=k*sin(thetay)
%
%c     with Kz=(cosx+cosy)*k we have
%c    gammamp = 1/9*Gammap ( 2+Cos[2*Kx*x]+Cos[2*Ky*y]-+4*Cos[Kx*x]*Cos[Ky*y]*Cos[Kz*z] )
\end{equation}
%where {\textcolor{black}{$k_x=k\sin\theta_x$,  $k$}} is the laser beam wave number, 
where {\textcolor{black}{$k_x=k_L\sin\theta_x$,  $k_L$}} is the laser beam wave number, 
$U_0=-16\hbar \Delta'_0/3 $, $\Delta'_0$ ($<0$) is the light-shift per lattice field,
and {\textcolor{black}{$\Gamma_S$}} is the photon scattering rate per lattice beam.
We have checked that, within the parameter range studied here, semiclassical simulations \cite{gryn98,scsims} of the 3D and 1D systems provide qualitatively equivalent results.
}
%\sout{Consider $F_{g} = 1/2 \leftrightarrow F_{e} = 3/2$ atoms confined in a 1D lattice formed by the superposition of two counterpropagating red-detuned laser beams 
%of amplitude 
%${\cal E}_0$,
%and frequency 
%$\omega$ of orthogonal linear polarization (lin$\perp$lin), as in Fig.~\ref{fig:three}(a). The lattice consists of light-shifted ground state $\pm 1/2$-spin potentials $U_\pm$, with wells $\lambda/4$ apart at sites of alternating circular polarization - see Fig. S1, supplementary notes.} 
{\dcor {\textcolor{black}{In each ground state,}} atoms
oscillate in optical wells with vibrational frequency
\begin{equation}
\Omega_X=k_x\sqrt{\frac{3 U_0}{2m}} = 4\sin\theta_x\sqrt{|\Delta_0^\prime|\,\omega_r},
\label{eq:omx}
\end{equation}
where $\omega_r= \hbar k_L^2/(2m)$ is the recoil frequency. But this motion is not uninterrupted, the interaction with the laser also produces}
%(curved arrows) 
%\sout{with mean oscillation speed $v_{osc}$, and undergo}
%are optically pumped by stochastic 
random absorption-emission processes 
%\sout{denoted by average photon scattering rate $\Gamma_s$ }
%(solid, dotted, dashed arrows) 
with the excited states (collectively denoted by $|e\rangle$). 
Photon scattering is viewed as noise, analogous to Brownian fluctuations in thermal systems, 
%because random recoils disrupt oscillatory motion in the wells,
%and damp 
%Hamiltonian motion in the reactive potentials, 
occasionally pumping an atom into {\dcor the other ground state sublevel at the rate {\textcolor{black}{$\gamma_{\pm}$ (\ref{eq:gammapm})}},
%a neighboring well at average rate $\Gamma_P = 4\Gamma_S/9$ (Eqn. S3, supplementary notes) 
and causing spatial diffusion~\cite{grynpra2003,palencia2002}. \textcolor{black}{The} transitions are most likely at the peaks of the potential barriers, a distance $\pi/k_x$ from the bottom of the well, yielding the well-known Sisyphus damping, where atoms mostly ``climb hills", continuously dissipating energy. }

%The same $\Omega_V$ can correspond to many $v_{osc}$-values, each mapping to a different oscillation amplitude in a classical picture.
%Atoms with low $v_{osc}$-values oscillate near the well-bottom, and likely remain in the well after photon scattering (dashed arrows).  
%Atoms located near the bottom of a well where the polarization is, say, pure $\sigma^{+}$ 
%For atoms with higher $v_{osc}$ the probability to be optically pumped to the adjacent well on either side grows significantly as the oscillation amplitude nears a $U_\pm$ crossing (solid vertical arrows), approaching unity as the atom nears the top of a hill (dotted arrows) - 
%this leads to Sisyphus cooling as is well-known.}
%this is the well-known Sisyphus damping, where the atom mostly ``climbs hills" and each successive well-transfer decreases $v_{osc}$. However, for the velocity-class corresponding to an oscillation amplitude equal to the $\lambda/4$-separation between adjacent $U_\pm$ crossings, the $v_{osc}$-value remains unchanged between well-transfers. These 
{\dcor Since the} atoms are most likely 
%to be 
pumped to a neighboring well {\dcor via sublevel transitions} at the turning points in their oscillations, 
{\dcor after half a {\textcolor{black}{time-}}period $\pi/\Omega_X$, the average velocity associated with this {\textcolor{black}{half-oscillation}} is given by $v_S= \Omega_X /k_x$. Once in the new ground state sublevel, it takes another half a period $\pi/\Omega_X$ to reach the 
%{\sout {new well bottom}} 
{\textcolor{black}{turning point on the other side of the well}}, also half a period length $\pi/k_x$ away. 
However, with no symmetry breaking element at play, each step can proceed with equal probability to the right or the left, yielding zero current. }

The introduction of a weak probe ${\cal E}_p$, $\omega_p$ {\textcolor{black}{propagating in the $z$-direction, as in}} {\dcor Fig.~\ref{fig:fig1}~(a),} breaks the symmetry resulting in directed motion {\textcolor{black}{along $\pm x$}}: ${\cal E}_p$ modulates the lattice at {\textcolor{black}{$\delta = \omega_L - \omega_p$}} 
{\dcor producing the following addition to the optical potentials
\begin{eqnarray}
U_\pm^p(x,t)=-2 U_0 \varepsilon_p \cos k_x x \cos \delta t \, ,
%U_\pm^p(x)=-\frac{U_0}{2}  \frac{{\cal E}_p}{{\cal E}_0} \cos(k_x x) \cos(\delta t ).
   \label{eq:probe:U}
\end{eqnarray}
where {\textcolor{black}{$\varepsilon_p={\cal E}_p/4{\cal E}_0$}}. 
%{\textcolor{black}{$\varepsilon_p={\cal E}_p/2{\cal E}_0$}}.
Since 
\begin{equation}
\cos k_x x \cos \delta t = \frac{1}{2}\Big[ \cos(-k_x x - \delta t) + \cos(+k_x x -\delta t) \Big],
\label{eq:cos:part}
\end{equation}
the probe contribution (\ref{eq:probe:U}) can be seen as the superposition of two perturbations traveling in opposite directions 
%{\textcolor{black}{$\Delta k_{1,2}$ in Fig.~\ref{fig:fig1}(a)}}, 
with velocity $\pm \delta/k_x$. Each perturbation is expected to excite an atomic density wave with wave number $k=\pm k_x$ and frequency $\omega=\delta$}, referred to as Brillouin modes in analogy to acoustic waves rippling through fluids~\cite{grynprl1996,grynprl2002}.  
{\dcor 
Optimal propagation is then expected when the velocity of these atomic modes, $v_B=\pm\delta/k_x$, matches  that of the {\textcolor{black}{half-oscillations}} discussed above, $v_S$, yielding a maximum at $\delta=\Omega_X$. Previous research  \cite{grynprl1996,grynprl2002,grynbergepl,grynpra2003} has focused only on these modes, which share the same frequency and wave number as the probe perturbations {\textcolor{black}{(a simple visualization is provided in Appendix C)}}.  The results of a novel theory \cite{cubero22}, presented in Fig.~\ref{fig:david}, do confirm them as the dominant ones, though other excited density waves can be clearly observed at play with non-negligible quantitative contributions. 

%%%%%%%%%%%%%%%%%%%
\begin{figure}[h]
\includegraphics[width=8cm]{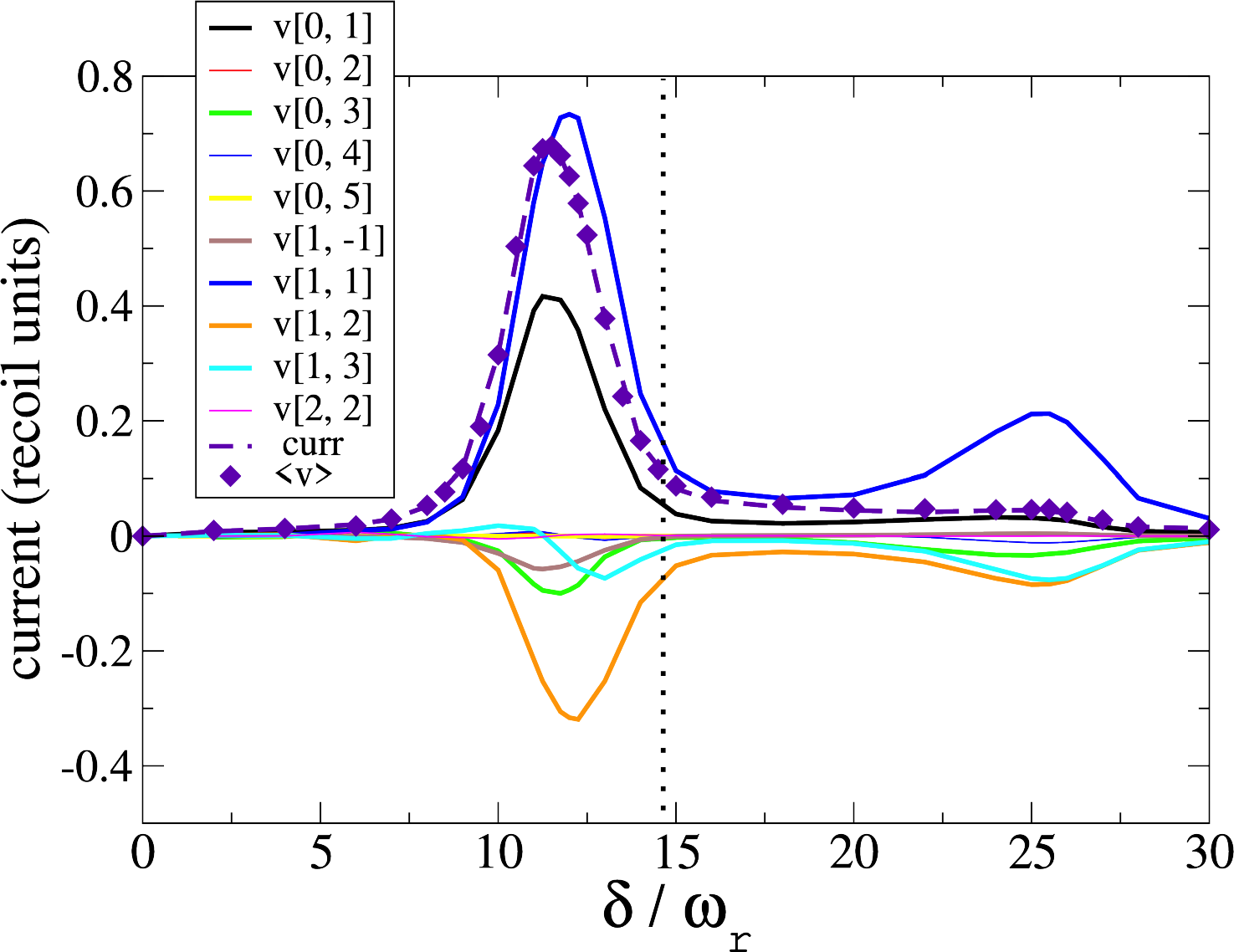}
%\vspace{-2mm}
\caption{\label{fig:david}
\dcor
Contribution to the current $v[{\textcolor{black}{l}},n]$ of the atomic density wave with frequency $\omega= {\textcolor{black}{l}} \delta$ and wave number $k=n k_x$  under a {\textcolor{black}{probe-induced lattice modulation traveling in the +x-direction}} (\ref{eq:probe:U:right}) with $\varepsilon_p=0.1$, from a semiclassical simulation with $U_0=400\,\hbar\omega_r$ and $\Gamma_S=5.7\,\omega_r$, $\theta_x=25^{0}$. The dashed line is the sum of all contributions, and the diamonds the current calculated directly in the simulation.  The vertical dotted line indicates the vibrational frequency $\Omega_X$. 
%% U0_thispaper=4*U0_sim,   thus U0_thispaper= 8*U0_sim* Er
%%  U0_thispaper/4*(Ep/E0) =U0_sim*EP, thus Ep/E0=EP
%% Gamma_s=GP_sim/2, thus Gamma_s=(1/2)*GP_sim/2/(1/2), thus Gamma_s/omegaR=GP_sim
%% U0_sim=50, GP_sim=5.7, EP=0.4,   Er=hbar*omegaR=hbar^2*kl^2/(2*m)=1/2
%% v(recoil)=v_sim/(2*pi/kl)*(2*pi/omegaR) = v_sim*kl/omegaR = v_sim*2
%% in sims hbar=kl=m=1.
}
%\vspace{-3mm}
\end{figure}
%%%%%%%%%%%%%%%%%%%

{\dcor In general, a propagating perturbation is expected to excite not only the atomic wave of frequency and wave number of the perturbation, but also other nearby modes. This is indeed the case in our system, Fig.~\ref{fig:david} shows that atomic waves with $\omega=\delta$ and $k=2k_x$ and $3k_x$ are {\textcolor{black}{also}} excited, as well as the mode $\omega=2\delta$, $k=2k_x$, which has the same phase velocity as the propagating perturbation, \textcolor{black}{in addition to}} a plethora of non-propagating modes (i.e. with $\omega=0$). 

{\textcolor{black}{However,}} mode excitation is not sufficient to guarantee a significant contribution to the directed motion. In Brownian ratchets ~\cite{cuberobook,hanggi1,reimann}, the quantity of interest is the current, defined as the average velocity $\langle v \rangle=\lim_{t\rightarrow\infty}[\langle x(t)\rangle-\langle x(0)\rangle]/t$. A novel theoretical development  \cite{cubero22} has been able to express analytically the current as an expansion $\langle v \rangle=\sum_{{\textcolor{black}{l}},n}v[{\textcolor{black}{l}},n]$, where $v[{\textcolor{black}{l}},n]$ is proportional to the Fourier amplitude of the atomic density wave with  frequency $\omega= {\textcolor{black}{l}} \delta$ and wave number $k=n k_x$. The analytical calculation is based on the coupled Fokker-Planck equations resulting from the semiclassical approximation \cite{scsims}. Fig.~\ref{fig:david} shows the results for the {\textcolor{black}{optical lattice}} (\ref{eq:Upm})--(\ref{eq:gammapm}) {\textcolor{black}{modulated by the $+x$-propagating part of the probe potential in (\ref{eq:probe:U})--(\ref{eq:cos:part})}}
\begin{eqnarray}
U_\pm^p(x)=- U_0 \varepsilon_p \cos(k_x x-\delta t ).
% U_\pm^p(x)=- \frac{U_0}{4}  \frac{{\cal E}_p}{{\cal E}_0} \cos(k_x x-\delta t ).
   \label{eq:probe:U:right}
\end{eqnarray}
The good agreement between the current {\textcolor{black}{obtained}} directly in the simulations (diamonds) and the sum of all the mode contributions (dashed line) serves as a validation of the analytical calculations  \cite{cubero22}. 

As discussed above, optimal transport for atoms following the density mode $\omega=\delta$, $k=k_x$ is expected when its propagation is synchronized with the most likely transitions between the ground state sublevels, the latter happening at half oscillations in the potential wells, and thus at $\delta\approx\Omega_X$. Here an atom could be hopping between states after half oscillations while riding at a maximum of the atomic wave. 
 Indeed, the strongest peak observed in Fig.~\ref{fig:david} takes place at a frequency slightly below $\Omega_X$. 
 {\textcolor{black}{The}} expression used for the vibrational frequency (\ref{eq:omx}) is based on a harmonic approximation for small deviations about the well's bottoms, actually underestimating the time spent in the half oscillation, and thus, overestimating the optimal frequency. 
 
 {\textcolor{black}{A }}second peak, though of considerably smaller amplitude, can be seen at double {\textcolor{black}{this optimal}} frequency in Fig.~\ref{fig:david}, indicating  a further synchronization mechanism between the two propagation processes. It can be also readily rationalized, as in these conditions, during a half oscillation time, the atomic wave, traveling at two times the previous speed, can still offer a density maximum at the same places where the atomic transitions are most likely, the potential barriers.
 
{\textcolor{black}{Note that Fig.~\ref{fig:david} shows}} important contributions to the directed motion coming from other modes. In particular, the atomic wave with $\omega=\delta$, $k=2k_x$ is observed to produce a current, slightly less than half the contribution of the probe's mode, in the opposite direction. In this case the atomic wave is moving with half the speed, $\delta/(2k_x)$, and has half the wavelength, $2\pi/(2k_x)$, thus there is an extra density maximum within the well length $2\pi/k_x$. Under these conditions, an atom moving in the direction of the probe-{\textcolor{black}{modulated potential (\ref{eq:probe:U:right})}} has higher speed than the wave and is less likely to cross a density maximum than when moving {\textcolor{black}{in}} the opposite direction, {\textcolor{black}{thus}} favoring that reversed direction. 
 
{\textcolor{black}{Moreover, Fig.~\ref{fig:david} shows that the contribution to the current from the mode with $\omega=0$, $k=k_x$ is slightly more than half the contribution of the dominant mode, and is therefore relevant.}} This kind of non-propagating mode is responsible {\textcolor{black}{for}} directed motion in rocking ratchets \cite{cuberobook}, where the driving force is usually chosen to be unbiased and non-propagating. Here the mode, like all modes shown in  Fig.~\ref{fig:david}, is also observed to be optimally excited at about the vibrational frequency, owing to the discussed matching {\textcolor{black}{of $\delta$}} with the intrinsic frequency {\textcolor{black}{$\Omega_X$}}.

 Note that the transition rates $\gamma_{\pm}$
 %(Eqn.~\ref{eq:gammapm}) 
 define {\textcolor{black}{another}} intrinsic frequency {\textcolor{black}{of the unperturbed lattice}}. Taking the spatial average of  (\ref{eq:gammapm}) results in $\gamma_0=2\Gamma_S/3$, the average number of atomic transitions per unit time.
 } 
Thus, a \emph{resonant enhancement} of atoms undergoing directed propagation {\dcor is expected} when $\gamma_0$ is synchronized with $\delta$ and $\Omega_X$. 
 {\dcor Specifically, the most coherent motion is expected when there is a single transition every half oscillation, $\gamma_0 (\pi/\Omega_X) \approx1$, yielding the  prediction~\cite{grynpra2003}
 \begin{equation}
 \Gamma_S  \approx {\textcolor{black}{\frac{6}{\pi}}}\sin\theta_x\sqrt{|\Delta_0^\prime|\omega_r}.
 \label{eq:gamsSR}
 \end{equation}
 }
Optimization of the unidirectional propagation {\dcor is therefore expected} when the random photon scattering process $\Gamma_S$ {\dcor is tuned to the value given by (\ref{eq:gamsSR}).} 
 This is stochastic resonance \cite{grynprl2002,grynbergepl,grynpra2003}, 
%\emph{Stochastic resonance is revealed directly by a peak in atomic current in the $+ z$-direction as $\Gamma_S$ is varied}.
 {\dcor which we demonstrate experimentally via pump-probe spectroscopy below.}

Our experiments are performed in a standard 3D tetrahedral lin$\perp$lin lattice, depicted in Fig.~\ref{fig:fig1}(a), comprising four equally intense near-resonant red-detuned beams (\textcolor{black}{$1/e^2$-diameter, 9.2 mm}; $\theta_x = \theta_y = 25^{0}$) that confine about $10^8$ $^{85}$Rb atoms ($\sim 30 \mu$K), i.e., only a few percent of the wells are occupied by an atom. We introduce a weak $y$-polarized $z$-propagating  probe beam (Z-probe) {\textcolor{black}{of intensity $I_p$}} with $1/e^2$-diameter 1.4 mm ($\lesssim$ diameter of cold atom cloud) at frequency $\omega_{p}$. 
%and angle $\theta_p$ with propagating along the $z$-axis, 
%that modulates the lattice at $\omega_{mod}$.
The lattice beams collectively serve as the pump at fixed frequency $\omega$. The probe frequency $\omega_{p}$ is scanned around $\omega$, and probe transmission is measured as a function of the pump-probe detuning $\delta$.
The intensity for a single lattice beam $I$ ranges from \textcolor{black}{1.48 to 14.5} mW/cm$^2$ (total lattice intensity $= 4I$), and the lattice is red-detuned by \textcolor{black}{$\Delta = 3.5 - 17$} $\Gamma$ from the $^{85}$Rb $F_g = 3 \rightarrow F_e = 4$  $D_2$  transition (natural linewidth $\Gamma/2\pi =$ \textcolor{black}{6.07 MHz}). {\dcor In all cases}  $\delta / \omega < 10^{-9}$ and 
the 
%%probe intensity is $< 1\%$ of the sum of lattice beam intensities. 
%{\textcolor{black}{probe-to-lattice intensity ratio $|\epsilon_p|^2$
{\textcolor{black}{intensity ratio of probe-to-single lattice beam is $|{\cal E}_p/{\cal E}_0|^2 < 4\%$, i.e., $I_p/4I < 1\%$, or $\epsilon_p < 0.05$.}}

\begin{figure}[t]
%%\vspace{-3mm}
%\centering
\includegraphics[width=8.6cm]{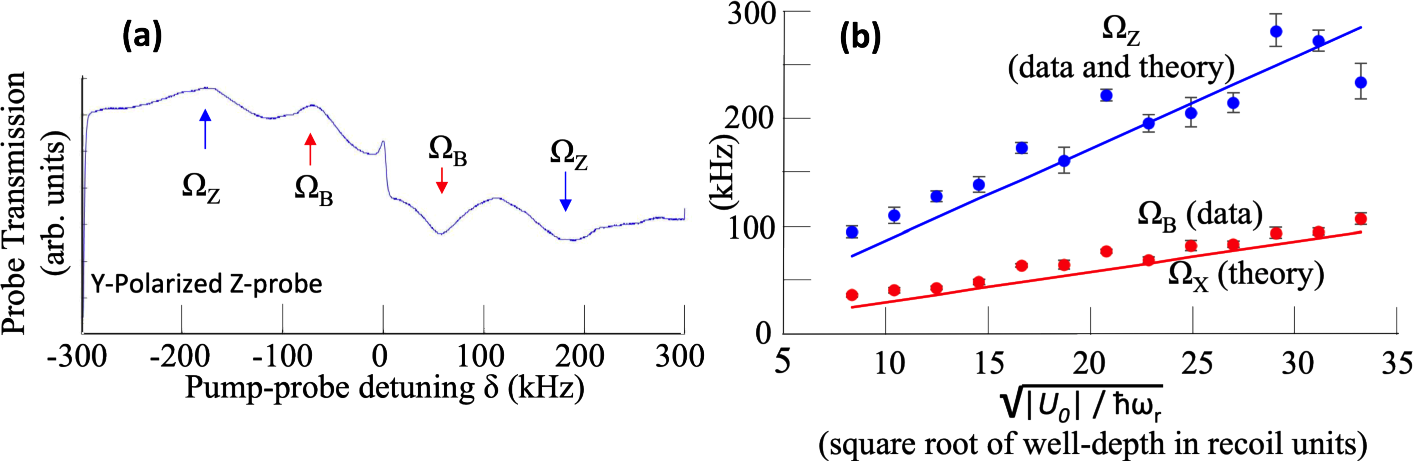}
%\vspace{-3mm}
\caption{{(a) 
Typical probe transmission spectrum for the 3D tetrahedral lin$\perp$lin lattice with weak $y$-polarized $z$-propagating probe, at well-depth $|U_0|/\hbar = 276\; \omega_r$ ($I = 5.2$ mW/cm$^2$, $\Delta = -12 \Gamma$; 10 ms for each scan). We observe a feature $\Omega_Z$ at \textcolor{black}{$\sim \pm 170$} kHz, and
%Really, $|U|/\hbar = 380\; \omega_r$ ($I = 6$ mW/cm$^2$, $\Delta = -10 \Gamma$
% \textcolor{black}{time for scan? plot?}
%The $z$-component $\Omega_Z$ of the vibrational frequency is observed at $\sim 170$ kHz. 
additional resonances $\Omega_B$ at $\sim \pm 60$ kHz which
%, which coincide with the $x$-component $\Omega_X$ of the vibrational frequency indicating 
arise from probe-induced directed atomic transport along $\pm \, x$.
%$\Delta \vec{k}_{1,2} = \vec{k}_{probe} - \vec{k}_{1,2}$ (see inset). %i.e., approximately along $- x$ and $+ x$, respectively (see text).
%lie mostly along $\pm x$; the small $z$-component is ignored, see text).
% when $\omega_{mod} \,(\,= \delta)$ coincides with $\Omega_X$. 
%\newline 
(b) Measured values for $\Omega_Z$ and $\Omega_B$ versus well-depth agree with calculations (lines) of the $z$-component of the vibrational frequency and $x$-component $\Omega_X$, respectively (see text). Here, $\omega_r$ is the $^{85}$Rb recoil frequency 3.86 kHz. No fitting parameters are used.
%  (see Eqns. S16, supplementary notes). 
 %No fitting parameters are used to draw the theoretical lines in (d).
 }}
%\vspace{-3mm}
\label{fig:spectraz}
\end{figure} 
%The 3D lattice is preferred to 1D because it confines atoms for longer times, yielding higher signal-to-noise ratio. 
%As shown in Fig.~\ref{fig:spectraz}(a), 
%We introduce a weak $y$-polarized $z$-propagating  probe beam (Z-probe) with $1/e^2$-diameter 1.4 mm ($\lesssim$ diameter of cold atom cloud) at frequency $\omega_{p}$. 
%%and angle $\theta_p$ with propagating along the $z$-axis, 
%%that modulates the lattice at $\omega_{mod}$.
%The lattice beams collectively serve as the pump at fixed frequency $\omega$. The probe frequency $\omega_{p}$ is scanned around $\omega$, and probe transmission is measured as a function of pump-probe detuning $\delta 
%\equiv \omega - \omega_{p}$.
%%(= \omega_{mod})$.
%%

To explore stochastic resonance we tune the \textcolor{black}{random} noise $\Gamma_S$ without changing 
% - dissipative optical lattices are suited for this.
%Dissipative optical lattices are suited for exploring stochastic resonance because this requires us to 
%%Dissipative optical lattices are useful for experimentally characterizing stochastic resonance which requires us to 
%%in an optical lattice 
%tune stochastic noise $\Gamma_S$ while leaving the system unchanged. 
the optical lattice, described by well depth $U_0$. From (\ref{eq:a1} - \ref{eq:a5}), we see that if we tune lattice intensity $I$ and detuning $\Delta$ such that the ratio $I/\Delta$ remains constant, we may vary $\Gamma_S$ ($\propto I/{\Delta}^2$) without changing $U_0$ ($\propto I/\Delta$).
%$U_0$ 
%%and vibrational frequency $\Omega_V$ are 
%varies as the ratio of lattice intensity to detuning $I/\Delta$, whereas 
%%the average photon scattering rate 
%$\Gamma_S$ varies as $I/{\Delta}^2$ (Eqns. S1-3, supplementary notes). If we tune $I$ and $\Delta$ such that 
%%$I/\Delta^2$ varies, thus tuning the stochastic noise, but 
%${I/\Delta}$ stays constant, we may vary $\Gamma_S$ without changing $U_0$.
%%, making dissipative optical lattices an ideal testbed for exploring stochastic resonance. 
 
Fig.~\ref{fig:spectraz}(a) shows the probe transmission spectrum 
%for $x$ and $y$ probe-polarizations, respectively, 
for a specific lattice well-depth.
%As explained below the additional spectral features in the $y$-polarized probe provide evidence for a spontaneous emission-enabled velocity-selective ratchet.
%Common to both plots is a 
\textcolor{black}{A spectral feature denoted $\Omega_Z$
%This feature's frequency-dependence on well depth is in good agreement with that of the calculated {$z$-component} $\Omega_Z$ of the vibrational frequency, 
%of \textcolor{black}{166 kHz}, 
%as shown in Fig.~\ref{fig:spectraz}(d).
%obtain similar pump-probe spectra for many different values of the well-depth (), we find that these observed $\Omega_Z$-values , as shown in Fig.~\ref{fig:spectraz}(d). 
arises because the Z-probe 
%``shakes" the lattice in the $z$-direction 
induces Raman transitions between adjacent vibrational levels in each well~\cite{grynprl1996}, in this case separated by $\sim \pm 170$ KHz. The peak (dip) 
%at $- (+) \, 170$ KHz 
corresponds to photons absorbed from a pump beam (the probe) and emitted into the probe (a pump beam).
%The features $\Omega_Z$ in Figs.~\ref{fig:spectraz}(b) and (c) %~\cite{grynbergepl,grynprl2002,grynprl1996}.
%, the key points are summarized in the supplementary notes~\cite{sup}. arise from 
%, thereby appearing as a gain (dip) in probe intensity. The dip at $+ 200$ KHz occurs due to photons that are absorbed from the probe and emitted into a pump mode.  (\textcolor{black}{further explanation supplementary notes?}). 
Further, we observe Brillouin resonances}
%at \textcolor{black}{$59 $} KHz, 
denoted as $\Omega_B$ which are a signature of directional transport as explained below. Fig.~\ref{fig:spectraz}(b) shows that the observed $\Omega_Z$ and $\Omega_B$-values, as the well-depth is varied, {\textcolor{black}{are in good agreement with the calculated values, shown by lines, for $\Omega_Z$ (\ref{eq:b4}) and $\Omega_X$ (\ref{eq:omx}), respectively.}}

%in Fig.~\ref{fig:spectraz}(d) 
%without using any fitting parameters.
%of \textcolor{black}{55} kHz. 
%Theoretical expressions for $\Omega_Z$ and $\Omega_X$ in a 3D tetrahedral lattice are well-known
 The features $\Omega_B$, 
 %in Fig.~\ref{fig:spectraz}(c), 
despite coinciding with $\pm \Omega_X$,
cannot arise from non-propagating atoms oscillating inside wells, because 
%the Z-probe can \emph{only} detect $\Omega_Z$. This is because 
%adjacent vibrational levels in a well have opposite parity, and 
the Z-probe operator is 
%linear in $z$ but 
quadratic in $x$ (Appendix D).
%~\cite{grynprl1996}. 
%Instead, there is breaking of symmetry leading to preferred propagation in the $x$-direction (\textcolor{black}{and y?}) owing to synchronization between $\omega_{mod}$ and $\Omega_X$, as briefly indicated in Fig.~\ref{fig:three}(a). 
% To see when synchronization  between $\omega_{mod}$ and $\Omega_X$ occurs we note that 
Instead, the 
%$\hat{y}$-polarized 
Z-probe interferes with the lattice beams, producing propagating modulations 
%along $\pm x$ and $\pm y$-directions 
that drive 
%Brillouin modes. 
directed transport.
%Our analysis 
%%below 
%is simplified by the fact 
Note that the
 contributions to the pump-probe spectrum 
 %in Fig.~\ref{fig:spectraz}(c) 
 from two of the four lattice beams,
 %from photon exchange between the probe and lattice beams 
 $\vec{k}_3$ and $\vec{k}_4$, are suppressed due to Doppler-broadening in the $z$-direction~\cite{aspect1998}.
 %$z$-component of atomic motion~\cite{aspect1998}, 
 %hence are ignored.

The interference of the 
$\hat{y}$-polarized 
probe with
%nearly co-propagating $\hat{y}$-polarized (counter-propagating $\hat{x}$-polarized) 
$\hat{y}$-polarized 
lattice beams $\vec{k}_1$, $\vec{k}_2$ ($|\vec{k}_{1, 2}| = |\vec{k}_p| = k$) 
generates a propagating
%is analogous to the 1D case of the probe polarized \emph{parallel} to the counter-propagating lattice beam mentioned above (Sec. 2.2 and 3.1, supplementary notes). 
\emph{intensity} modulation 
%in analogy to the 1D model, 
in directions $\pm x$ 
{\dcor (\ref{eq:probe:U}-\ref{eq:cos:part}).
%$\Delta \vec{k}_{1}$ and $\Delta \vec{k}_{2}$ (inset, Fig.~\ref{fig:spectraz}a), which align approximately with the $-x$ and $+x$ axes, at speeds $\pm \, \delta/(k \, \mbox{sin} \, \theta_x)$. The small $v_{mod, \, z}$ component can be safely ignored because $\Gamma_\pm$ vary quadratically along $z$, in contrast to quartic along $x$ (Eqn. S9, supplementary notes) making well-to-well transfer via optical pumping along $z$ inefficient.
%Thus,
%(k \, \mbox{sin} \, \theta_x)
%the velocity-matching condition necessary for 
As discussed above, optimal }
directed propagation is expected at  
%The $x$-component of the intrawell oscillation speed 
%$v_B = v_{mod, \, x}$ where $v_B = \Omega_X / (k \, \mbox{sin} \, \theta_x)$, which yields   
$\delta = \pm \Omega_X$, precisely the features marked $\Omega_B$ in Fig.~\ref{fig:spectraz}(a). This velocity class, instead of diffusing in all directions, is ratcheted
% by the modulation 
%as in Fig.~\ref{fig:three}(d) 
along $\pm x$, 
%not just $+x$, 
yielding a bidirectional ratchet. 

\begin{figure}[b]
%%\vspace{-5mm}
%\centering
\includegraphics[width=8.7cm]{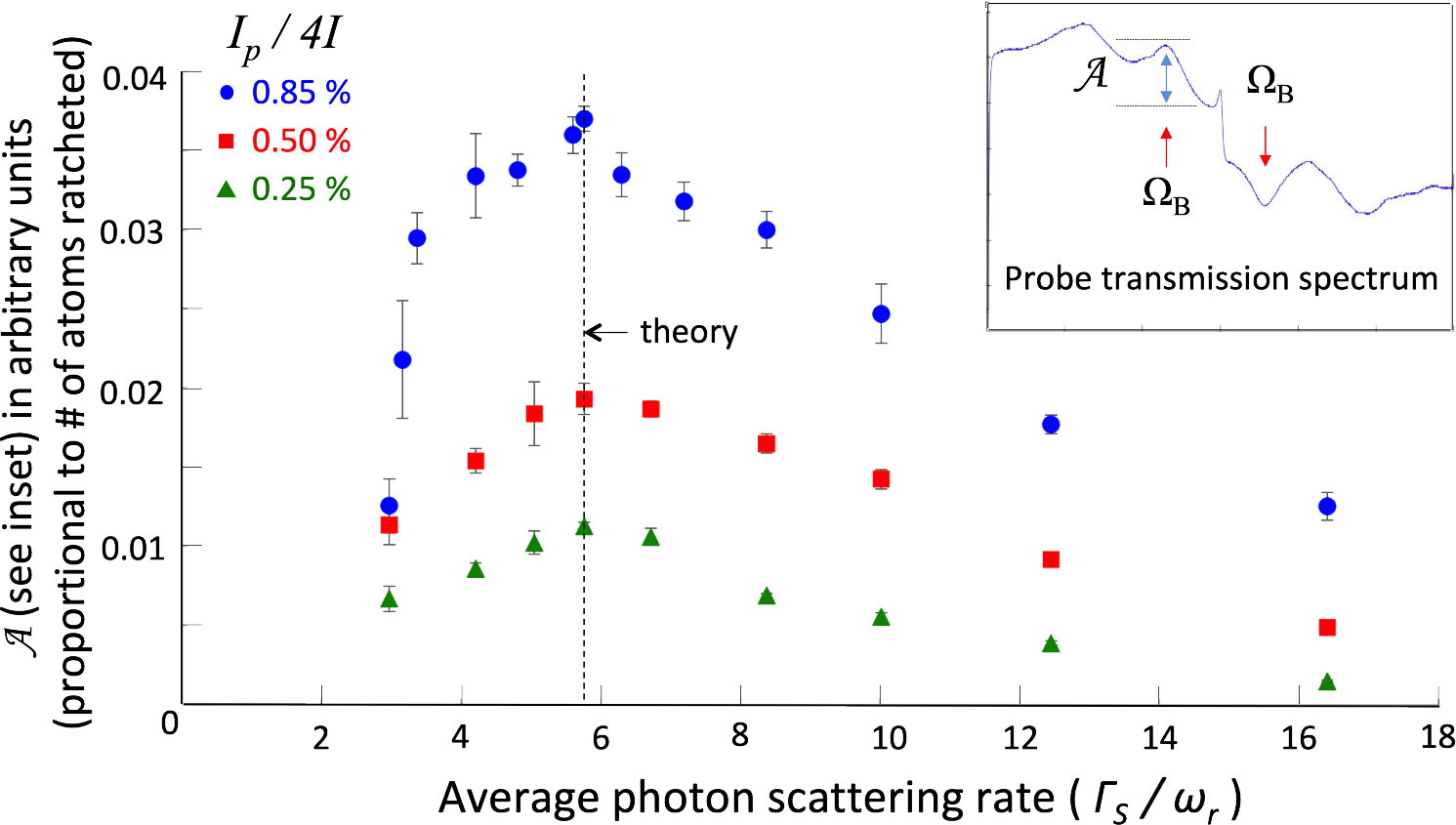}
%\vspace{-2mm}
\caption{{{Number of atoms ratcheted along $\pm x$ versus noise rate, or more specifically: Amplitude of Brillouin peak $\Omega_B$ vs. $\Gamma_S$ the photon scattering rate per lattice beam in units of recoil frequency $\omega_r$, at a fixed well-depth ($|U_0|/\hbar = 261 \omega_r$). Stochastic resonance is observed at $\Gamma_S/\omega_r = 5.7$ irrespective of modulation amplitude, in agreement with theory (\ref{eq:gamsSR}). 
%raph showing the effect of modulation amplitude on the stochastic resonance curve. We see that we can use modulation amplitude as a way to optimize stochastic resonance and achieve a greater efficiency for our nanoratchet.   
}}}
%\vspace{-2mm}
\label{fig:SR_1}
\end{figure}

\begin{figure}[t]
%%\vspace{-4mm}
%\centering
\includegraphics[width=8cm]{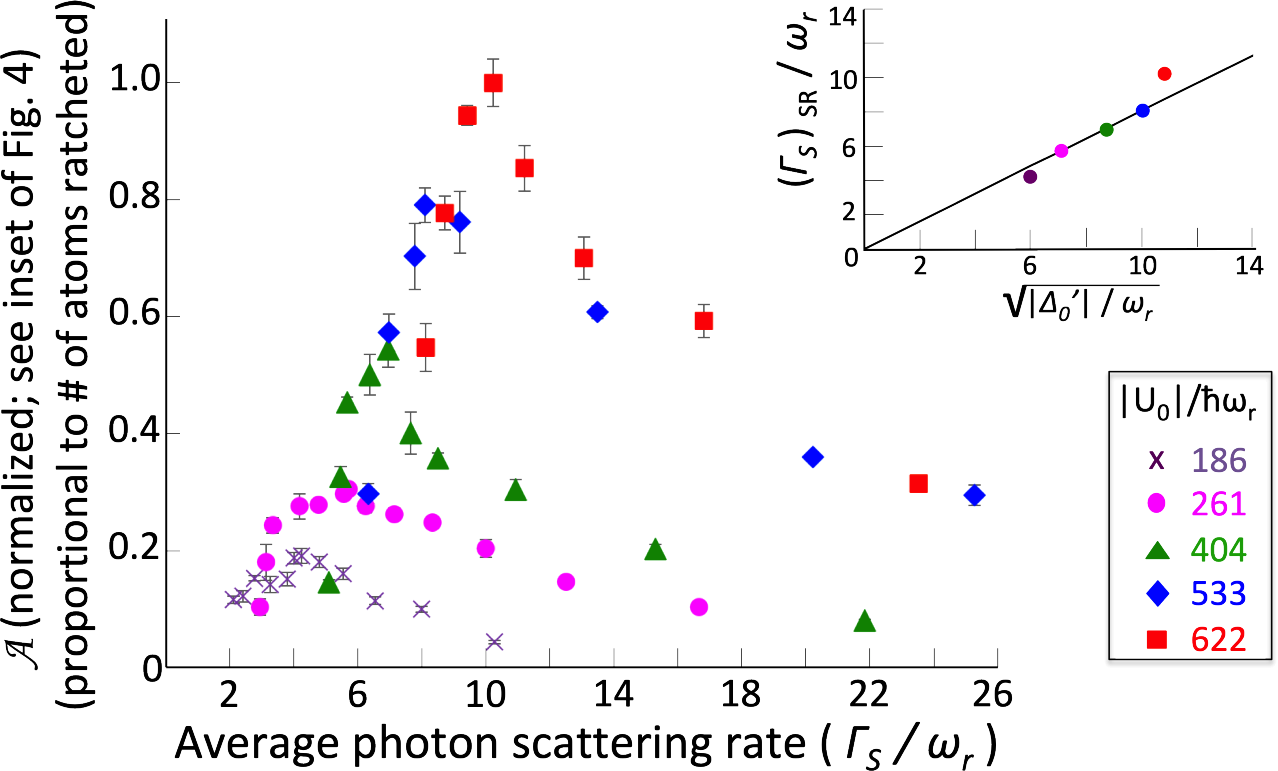}
%%\vspace{-6mm}
\caption{{{Number of atoms ratcheted along $\pm x$ vs. noise rate, for different well-depths $|U_0|/\hbar \omega_r$ at fixed modulation strength $I_p/4I = 0.85\%$. Inset: The noise rate $({\Gamma_S})_{SR}$ at which stochastic resonance occurs scales linearly with $\sqrt{|{\Delta_0}'|}$, in good agreement with a theoretical line drawn using (\ref{eq:gamsSR}) with no fitting parameters. 
%Here, $U_0 = \frac{16}{3} \hbar {|\Delta_0}'|$.
%c) The relative magnitude of the resonantly enhanced directed current ($\propto {\cal A}_{SR}$) is plotted versus the light shift ${\Delta_0}'$. 
%raph showing the effect of modulation amplitude on the stochastic resonance curve. We see that we can use modulation amplitude as a way to optimize stochastic resonance and achieve a greater efficiency for our nanoratchet.   
}}}
%\vspace{-5mm}
\label{fig:SR_2}
\end{figure}

In a significant departure from previous works, we shift attention from the 
location of the 
peak $\Omega_B$ 
%for directed transport 
to the \emph{peak-amplitude} $\cal A$ which is proportional to the \emph{number of atoms ratcheted} along $\pm x$. Figs.~\ref{fig:SR_1} and~\ref{fig:SR_2} show measurements of $\cal A$ in the probe-modulated 3D lattice as a function of the stochastic noise 
%frequency represented by the average photon scattering 
rate $\Gamma_S$. 
%graphs of the efficiency of directed propagation along $\pm x$-directions in the probe-modulated 3D lattice
Each data point is an average $\cal A$-value measured from \textcolor{black}{at least five} scans similar to Fig.~\ref{fig:spectraz} (a) (see inset, Fig.~\ref{fig:SR_1}),
% at fixed  $I$ and $\Delta$. 
%lattice intensity and detuning. 
using a curve-fitting procedure 
%where the pump-probe spectrum is fitted with a sum of four Gaussian functions, two each for the $\Omega_Z$ and $\Omega_B$ peak/dip features 
(Appendix E). 
%$\Gamma_S$ is varied by tuning $I$ and $\Delta$, keeping $I/\Delta$ constant.
%so as to leave the lattice itself unchanged, as explained earlier. 
If the number of atoms initially confined in the lattice is {\textcolor{black}{kept the same}} (\textcolor{black}{within 7\% in our case}), and $I/\Delta$ is held constant as we vary $\Gamma_S$, the $\cal A$-values are a measure of \textcolor{black}{the directed atomic current} for different noise rates.
%, and can be compared.

%Our signal-to-noise 
%%in Fig.~\ref{fig:spectraz}(c) exceeds that of ENS data~\cite{grynprl96,grynprl}, 
%is sufficient to permit a search for stochastic resonance via pump-probe spectroscopy which is far more rapid than diffusion measurement.  
%, while the amplitude of the vibrational resonance $\Omega_Z$ is proportional to the number of atoms trapped inside potential wells. Thus, a comparison of peak-amplitudes of $\Omega_B$ versus $\Omega_Z$ would indicate the fraction of atoms participating in Brillouin propagation versus atoms sitting inside wells. 

%Otherwise, unwanted contributions from resonant activation, a rival process to stochastic resonance, may appear as discussed in section~\emph{\textcolor{blue}{\ref{subsec-backgr12}3}}. 
Fig.~\ref{fig:SR_1} shows clear evidence for 
%\textcolor{black}{spontaneous emission mediated} 
stochastic resonance in a modulated cold atom optical lattice for three {\textcolor{black}{probe modulation strengths $I_p/4I$
ranging from 
%a $I_p$-value
%3.4\% of the intensity $I$ of a single lattice beam (i.e., 
$0.25\%$ to 0.85\%.}} Increasing the modulation strength results in larger atomic current, but since the lattice, and therefore vibrational frequency, remains unchanged we observe stochastic resonance at the same $\Gamma_S$-value.
% for the different modulation strengths. 
 Atomic transport in this bidirectional ratchet falls off on either side of stochastic resonance, faster for low $\Gamma_S$-values since well-to-well transfer is disrupted more effectively if optical pumping simply does not occur. 
%which cause a larger disruption in the well-to-well transfer at the crossing points.
 %As expected, increasing the modulation amplitude results in a larger number of atoms dragged along.
%by the moving modulation.
%see Eqn. S2 of supplementary notes). 
These observations concur with numerical simulations in Ref.~\cite{grynpra2003} of the diffusion coefficient predicting enhanced transport along $\pm x$.
%obtained for the first time by pump-probe spectroscopy. 
%We infer from the preceding discussion that stochastic resonance in a 3D lattice with $z$-propagating probe occurs when two conditions are satisfied: First, the modulation frequency $\delta$ equals the $x$-component $\Omega_X$ of the oscillation frequency, yielding the Brillouin resonances $\Omega_B$ in Fig.~\ref{fig:spectraz}(c). Second, the photon scattering rate $\Gamma_S$. 

{\dcor Furthermore, the  noise rate at which stochastic resonance occurs is seen in Fig.~\ref{fig:SR_1} to be in very good agreement with the prediction of (\ref{eq:gamsSR}).
%The noise rate at which stochastic resonance occurs can be predicted by quantitatively expressing the situation depicted in Fig.~\ref{fig:three}(d): We equate the typical time between two subsequent optical pumping cycles $1/\overline{\Gamma_{\pm}} = \frac{9}{4{\Gamma_S}} \, (\overline{ {{\mbox{cos}}^2} {K_x}x + {{\mbox{cos}}^2}{K_y}y })^{-1}$ with that for half-oscillations in the $x$-direction $\pi/\Omega_X = \pi (4 \, \mbox{sin} \, \theta_x{\sqrt{|{\Delta_0}'|\omega_r}})^{-1}$ (Eqns. S8 - S9, supplementary notes), arriving at the prediction~\cite{grynpra2003}:
%%\vspace{-2mm}
%\begin{equation}
%(\Gamma_S)_{SR} = \frac{9 \, \mbox{sin} \, \theta_x \, \sqrt{|{\Delta_0}'|\omega_r}}{ \pi (\overline{ {{\mbox{cos}}^2} {K_x}x + {{\mbox{cos}}^2}{K_y}y }) }
%\label{eq:SRcondition}
%%\vspace{-2mm}
%\end{equation}
%\noindent 
%Eqn.~\ref{eq:SRcondition}, in addition to the aforementioned Brillouin condition $\delta = \pm \Omega_X$, achieves synchronization between $\Gamma_S$, $\Omega_X$, and $\omega_{mod} \, (= \delta)$. 
This agreement}
% in Fig.~\ref{fig:SR_1} between theory and experiment 
%for the $(\Gamma_S)_{SR}$-value 
is remarkable considering that the prediction uses a $F_{g}=1/2 \rightarrow F_{e}=3/2$ atom. 
%\textcolor{black}{shape of SR-curve?}
In Fig.~\ref{fig:SR_2} we test the theory further by fixing the modulation strength and investigating stochastic resonance for several different well-depths. {\textcolor{black}{In the inset we denote the noise rate at which stochastic resonance occurs as $(\Gamma_S)_{SR}$, and experimentally verify the linear dependence of $(\Gamma_S)_{SR}$ on the square root of the light shift $\Delta_0'$.
%The inset shows our experimental verification of the linear dependence {\textcolor{black}{on the square root of the light shift $\Delta_0'$ predicted in (\ref{eq:gamsSR}) for}} {\dcor the optimal noise value, which we denote by $(\Gamma_S)_{SR}$.} 
The requirement to keep the modulation strength fixed while varying well-depth ($\propto I$) means that
at the highest well-depth the probe intensity is strong enough to perturb the lattice, while at the lowest well-depth $I_p$ is weak and barely excites Brillouin propagation, causing the data to depart from theory in both cases. }}

In conclusion, we have demonstrated stochastic resonance in a modulated dissipative optical lattice, as a function of the noise rate, the modulation amplitude, and the lattice well depth. By observing the transmission spectrum of a weak probe beam that modulates the lattice we present evidence that the photon scattering rate at which stochastic resonance occurs is independent of the modulation strength, and that the stochastic resonance can be controlled by varying the lattice well depth. 
%By observing the transmission spectrum of a weak probe beam that modulates the lattice we have performed a first experimental investigation of the dependence of stochastic resonance on probe intensity and lattice well depth. 
Remarkably, the data agrees well with theory based on a simple $F_{g}=1/2 \rightarrow F_{e}=3/2$ atom without use of any fitting parameters. 
%A 1D model is used to elucidate how probe-induced modifications of the light-shifted ground state lattice potentials on one hand and the optical pumping rates on the other hand conspire to create resonant directional propagation within a randomly diffusing cold atom cloud. \textcolor{black}{Future? angled probe - mech res vs vel match}
{\dcor Furthermore, a recent novel theory has permitted us to precisely determine the contribution to the directed motion of the atomic density waves excited by the perturbing probe, and how they conspire with}
the optical pumping rates 
to create resonant directed propagation within a randomly diffusing cold atom cloud. \textcolor{black}{We believe that this work may contribute toward the quest for artificial nano devices that can operate in noisy environments with high efficiency~\cite{pekola2015,nat2019,nat2021}. 
A recent observation of stochastic resonance in quantum tunneling of an electron~\cite{timo,ludwig} reveals new directions where the source of the random noise is intrinsic quantum fluctuations in the system, instead of fluctuations in the environment. }

\section{Acknowledgments}
%%\vspace{-4mm}
This work is supported by the Army Research Office under award/contract number W911NF2110120.  {\dcor DC acknowledges financial support from the Ministerio de Ciencia e Innovaci\'on of Spain of Spain, Grant No. PID2019-105316GB-I00.} We thank the Instrumentation Laboratory at Miami University for electronics and LabView support. 
We thank A. Dharmasiri and A. Rapp for assistance during the initial setup. 
We are grateful to A. Reinhard and N. Peters for providing invaluable feedback on the manuscript.
%\end{acknowledgments}

%\Appendix
\section{Appendices}
Here, we provide additional proofs and
%mathematical proofs, pictorial justifications, and 
justifications for statements made in the main text, and summarize some relevant background on 
%one-dimensional and three-dimensional 
dissipative optical lattices.
%: Sec.~\ref{sec:1dlat} starts with the conceptually simpler one-dimensional configuration, while Section~\ref{sec:3dlat} focuses on the three-dimensional setup used in the reported experiments. 
%\textcolor{black}{In Section~\ref{sec:1dmodlat} we provide a simplified visualization of the dominant [1,1] Brillouin mode that propagates at the same frequency and wave number as the probe perturbation.} Section~\ref{sec:omb_omx} discusses the association of $\Omega_B$ in Fig. 3 with a propagating mode. In Section~\ref{sec:curvefit} we describe the fit function used to model our measured probe transmission spectra. 

%\renewcommand{\thesection}{A\arabic{section}}
\section{{
%$F_{g}=1/2 \rightarrow F_{e}=3/2$ atom in 
A: The 1D dissipative optical lattice}}
%: \newline how probe-modulation yields unidirectional motion}
\label{sec:1dlat}
\vspace{-2mm}
%\label{sec:disslat}
%The cartoon in Fig.~\ref{fig:stochres_cartoon} is suited to the picture of an atom rolling inside a potential well and occasionally hopping between adjacent wells, as in an optical lattice. In fact, 
%\vspace{-3mm}
Dissipative optical lattices are an ideal testbed for studying stochastic resonance because the stochastic coupling between the system (confined atom) and environment (random fluctuations in energy in the form of 
photon scattering) can be precisely controlled by varying the laser intensity and detuning, while ensuring that the lattice well-depth remains constant. This essential point can be conveyed using well-known results for a 1D lin$\perp$lin lattice and a $F_{g}=1/2 \rightarrow F_{e}=3/2$ atom.
% where $F$ is the hyperfine quantum number. 

%Optical lattices are cold atoms trapped in a crystal-like periodic structure of potential wells induced by the interference of several laser beams. 
%%Lattice well-depths and inter-well spacings can be precisely varied by adjusting the laser frequencies and intensities and the relative angles between the superposed laser beams. 
%In a ``dissipative" lattice the beams are red-detuned but near-resonance so that the atoms continually undergo random kicks owing to spontaneous emission, identical to thermal collisions in a bio-molecular motor.

%\vspace{-2mm}
%\subsubsection{$F_{g}=1/2 \rightarrow F_{e}=3/2$ atom in 1D dissipative optical lattice}
%\label{sec:1d}
Two red-detuned laser beams
%$\vec{E}_{1}$ and $\vec{E}_{2}$ 
of orthogonal linear polarization but same amplitude 
 %$\omega_{L}$ 
 and  
 %$E_{0}$ 
wavelength $\lambda$, counterpropagate along the $z$-axis, 
%as in Fig.~\ref{fig:1dlat}(a), 
yielding constant intensity but with steep polarization gradient of pitch
%cycling from linear to circular (say $\sigma^{-}$) to orthogonal-linear to opposite-circular ($\sigma^{+}$) to %original-linear, all within a distance of 
$\lambda$/2.
The two ground state magnetic sub-levels $m_{F_{g}} = \pm 1/2$ undergo polarization-dependent AC Stark shifts yielding spatially modulated potential wells 
%\textcolor{black}{$U_{\pm}(z)$} 
which have maximum depth \textcolor{black}{$U_{1}$} 
at sites of alternating pure circular polarization. Atoms settle at the bottoms of these wells. Their
%, shown in Fig.~\ref{fig:1dlat}(c). 
%Each atom is Sisyphus-cooled, eventually settling in a well where it oscillates with frequency $\Omega_V$. The 
oscillatory excursions away from sites of pure circular polarization cause optical pumping to adjacent wells on either side, leading to diffusion along $\pm z$.

In the weak excitation limit, we obtain the following well-known expressions~\cite{grynreview,cct,gryn1994,grynpra2003}
%the saturation parameter $s_0$, defined by the expression $\frac{\Omega^2/2}{\Delta^2+\Gamma^2/4}$ is always $<< 1$, i.e, the atom spends almost all its time in the two ground magnetic substates $|1, 2\rangle \equiv |F_g = 1/2, m_g = \pm 1/2\rangle$.
for the light-shifted bi-potential $U_{\pm}(z)$:
\setcounter{equation}{0}
\renewcommand{\theequation}{A\thesection\arabic{equation}}
\begin{equation}
U_{\pm} (z) = \textcolor{black}{\frac{U_1}{2}} \left[ -2 \, \textcolor{black}{\pm} \, \mbox{cos} \, 2k_Lz \right], 
\label{eq:a1}
\end{equation}
the 1D lattice well-depth $U_1$:
\begin{equation}
U_1  =  -\frac{4}{3}{\hbar {\Delta_0}'} 
 =  \textcolor{black}{-}\frac{2}{3} \left( \frac{I/I_{sat}}{1 + 4\Delta^2/\Gamma^2} \right) {\hbar \Delta} \propto \frac{I}{\textcolor{black}{|\Delta|}},
 \label{eq:a2}
 \end{equation}
the intra-well vibrational frequency $\Omega_V$:
\begin{equation}
\Omega_V = 2 \sqrt {\omega_{r} \, {U_1}/\hbar} \propto \sqrt{\frac{I}{|\Delta|}},
\label{eq:a3}
\end{equation}
and the position-dependent 
transition probability rates $\gamma_\pm$ between the ground state potentials: 
%optical pumping rate from $|1\rangle$ to $|2\rangle$ denoted by $\Gamma_{+}$, and from $|2\rangle$ to $|1\rangle$ denoted by $\Gamma_{-}$, are expressed 
 %(Fig.~\ref{fig:well})
%\vspace{-2mm}
%\renewcommand{\theequation}{S\arabic{equation}}
%\begin{eqnarray}
%%U_{\pm} (z) & = & \textcolor{black}{\frac{U_{1}}{2}} \left( -2 \mp \mbox{cos} \,2kz \right),
%U_{\pm} (z) & = & \textcolor{black}{\frac{U_1}{2}} \left[ -2 \, \textcolor{black}{\pm} \, \mbox{cos} \, 2k_Lz \right], \, \, \mbox{where}\nonumber \\ 
%& & U_1 = \mbox{1D lattice well-depth in Fig. S1(c)} = -\frac{4}{3}{\hbar {\Delta_0}'} 
% =  \textcolor{black}{-}\frac{2}{3} \left( \frac{I/I_{sat}}{1 + 4\Delta^2/\Gamma^2} \right) {\hbar \Delta} \propto \frac{I}{\textcolor{black}{|\Delta|}} \nonumber \\
% \label{eq:s1}
% & & \\
%\label{eq:s2}
%\Omega_V & = & 2 \sqrt {\omega_{r} \, \textcolor{black}{U_1}/\hbar} \propto \sqrt{\frac{I}{\textcolor{black}{|\Delta|}}}  \\
% \label{eq:s3}
\begin{equation}
\gamma_{\pm} (z) = \frac{2\Gamma_S}{9} \, ({1 \pm \mbox{cos} \, 2k_Lz }), 
\label{eq:a4}
\end{equation}
where
\begin{equation}\Gamma_S = \frac{\Gamma}{2} \left( \frac{I/I_{sat}}{1 + 4\Delta^2/\Gamma^2} \right) \propto \frac{I}{{\Delta}^2}, 
% SEE PAGE 45 OF NOTEBOOK "OPTICAL LATTICES -2 "
\label{eq:a5}
\end{equation}
\noindent Here $I$ is the lattice laser intensity (per beam) and  $I_{sat}$ is the saturation intensity (1.67 mW/cm$^2$ for the $^{85}$Rb D$_2$ $F_g = 3, m_{Fg} = \pm 3\rangle \rightarrow |F_e = 4, m_{Fe} = \pm 4\rangle$ $\sigma^{\pm}$ cycling transitions), $\Delta$ is the lattice laser detuning 
%(defined positive for red detuning), 
\textcolor{black}{($< 0$)}, 
$\Gamma/2\pi$ is the natural linewidth and $\omega_r/2\pi$ is the recoil frequency (6.07 MHz and 3.86 kHz, respectively, for $^{85}$Rb), $k_L = 2\pi/\lambda$,
%where $\lambda$ is the laser wavelength, 
and $z$ is the lattice axis. The relation $\textcolor{black}{|\Delta|} >> \Gamma$ usually applies in practical situations.
%, even if approximately. 
The ratio $(I/I_{sat})/(1 + 4\Delta^2/\Gamma^2)$ is just the saturation parameter $s_0$ which is $<< 1$ in the weak excitation limit. In our 3D lattice experiments $s_0$ ranges from about 0.001 to about 0.08. 
%The symbol $\Gamma_p \equiv 2{\Gamma s_0}/9$ denotes an averaged optical pumping rate between the ground states, and 
${\Delta_0}' \equiv \Delta {s_0}/2$ is the light-shift per lattice beam for a closed transition having a Clebsch - Gordan coefficient equal to 1, in accordance with notation used in Refs.~\cite{grynreview} and~\cite{grynpra2003}. Finally, ${\Gamma_S} \equiv \Gamma {s_0}/2$ is the photon scattering rate per lattice beam~\cite{grynpra2003}.

%Dissipative optical lattices offer an ideal testbed for studying stochastic resonance because
%from Eqns.~\ref{eq:scatt}-\ref{eq:well}, 
From Eqns.~\ref{eq:a1} - \ref{eq:a5} we see that if we tune $I$ and $\Delta$ such that $I/\Delta^2$ varies but ${I/\textcolor{black}{|\Delta|}}$ stays constant, we can tune the stochastic noise given by the photon scattering rate $\Gamma_S$ while keeping the lattice unchanged (because the well-depth $U_1$ and vibrational frequency $\Omega_V$ stay constant). 
%This distinctly different scaling behavior means that the characteristic frequency of the system $\Omega_V$ is independently tunable of the frequency of environmental noise $\Gamma_S$, making the optical lattice an ideal and versatile architecture for studying stochastic resonance. Further, each of these frequencies can be tuned simply by adjusting the lattice laser intensity and detuning. 
%Thus  $\Gamma_S$ varies, while $U_0$ and $\Omega_V$ and hence the lattice stay unchanged. 
%(the third frequency in the mix, i.e., the modulation frequency $\omega_d$ - see Sec.~{\emph {\ref{subsec:clqusr}}} - the experimenter has freedom to choose). 
This ability to tune the stochastic noise while leaving the lattice unchanged makes the dissipative lattice a uniquely ideal testbed for investigating stochastic resonance. 
%as we shall see in \emph{\textcolor{blue}{Sec.~\ref{subsec-plan1}1}} 
%Finally, as is well known, lattice well depths and spacings can be precisely varied over a wide range of values simply by adjusting the laser frequencies and intensities and the relative angles between the superposed laser beams.

%Expressions for 3D lattice quantities such as the vibrational frequency, well-depth, and photon scattering rate, that are important for our experiments, are stated below.    

\section{
%$F_{g}=1/2 \rightarrow F_{e}=3/2$ atom in 
B: The 3D dissipative tetrahedral optical lattice}
\label{sec:3dlat}
Our experiments on stochastic resonance are carried out in a three-dimensional standard lin$\perp$lin dissipative ``bright" optical lattice. A standard 3D tetrahedral lin$\perp$lin lattice consists of four overlapping equally intense near-resonant red-detuned beams as shown in Fig.~\ref{fig:fig1}(a). 
%(same as Fig. 1a in the main paper, but without the probe beam). 
The total electric field now takes the form~\cite{grynreview}:
\setcounter{equation}{0}
\renewcommand{\theequation}{B\thesection\arabic{equation}}
\begin{eqnarray}
\vec{E} (\vec{r},t)  & = & \frac{1}{2} {{\cal E}_0} [ e^{i\phi} \hat{e}_y ( e^{i\vec{k}_1.\vec{r}} + e^{i\vec{k}_2.\vec{r}} ) \nonumber \\
 & &  + \, \, \, \, \, \, \, \,   \hat{e}_x ( e^{i\vec{k}_3.\vec{r}} + \hat{e}_x \, e^{i\vec{k}_4.\vec{r}} )] e^{-i\omega_L t} + c.c. \nonumber
%\nonumber \\
%& + & \frac{1}{2}{{\cal E}_p} \,  \, \hat{e}_{y} \, e^{ik_{p}z} \, e^{-i\omega_{p\parallel} t} \nonumber \\
%& + & c.c.
%\label{eq:E}
\end{eqnarray}
Here
\begin{eqnarray}
\vec{k}_1 \cdot \vec{r} & = & k_L (x\, \mbox{sin}\theta_x + z\, \mbox{cos}\theta_x) \nonumber \\
\vec{k}_2 \cdot \vec{r} & = & k_L (-x\, \mbox{sin}\theta_x + z\, \mbox{cos}\theta_x) \nonumber \\
\vec{k}_3 \cdot \vec{r} & = & k_L (x\, \mbox{sin}\theta_y - z\, \mbox{cos}\theta_y) \nonumber \\
\vec{k}_4 \cdot \vec{r} & = & k_L (-x\, \mbox{sin}\theta_y - z\, \mbox{cos}\theta_y) \nonumber
%\label{eq:kr}
\end{eqnarray}
\noindent where we make the choice $e^{i\phi} = -i$, as in Ref.~\cite{grynreview}.
%and the same choice of phase $\phi$ is made as in Ref.~\cite{grynreview}~\cite{origin}. 

For a $F_{g}=1/2 \rightarrow F_{e}=3/2$ atom confined in this 3D lattice the 
%\newline 
equations corresponding to Eqns.~\ref{eq:a1} - \ref{eq:a3} are derived in Refs.~\cite{grynreview,gryn1994,grynpra2003}:
\begin{eqnarray}
U_{\pm} (x, y, z) & = & \frac{U_0}{4} [{-2} -{\mbox{cos}}(2k_x x) - {\mbox{cos}}(2k_y y) \nonumber \\
& & \pm \, 2\, {\mbox{cos}(k_x x)} \,{\mbox{cos}(k_y y)} \,{\mbox{cos}(k_z z)}]
\label{eq:3dpot}
\end{eqnarray}
The $y = z = 0$ section of the potentials in (\ref{eq:3dpot}) is used in (\ref{eq:Upm}). Here, the 3D lattice well-depth $U_0$ is given by
\begin{equation}
U_0 =  -16\hbar{\Delta'_0}/3 .
%\frac{|U_{0}|}{\hbar \Delta} = \frac{2}{3} \left( \frac{I/I_S}{1 + 4\Delta^2/\Gamma^2} \right) \propto \frac{I}{\Delta} \\
\label{eq:s5}
\end{equation}
The vibrational frequencies and position-dependent transition probability rates are given by
\begin{eqnarray}
\Omega_{X,Y} & = & 4 \, \mbox{sin}\,\theta_{x,y} \sqrt {|{\Delta'_0}| \, \omega_{r}} \\
\label{eq:s6}
\Omega_Z & = & ( \mbox{cos}{\theta_x} + \mbox{cos}{\theta_y}) \sqrt {|\Delta'| \, \omega_r}  \\
\label{eq:b4}
\gamma_{\pm} (x, y, z) & = & \frac{2\Gamma_S}{9}  [ 2 + {\mbox{cos}}(2k_x x) + {\mbox{cos}}(2k_y y) \nonumber \\
& & \pm \, 4\,{\mbox{cos}(k_x x)} \,{\mbox{cos}(k_y y)} \,{\mbox{cos}(k_z z)} ]
%\label{eq:s6}
\end{eqnarray}
% SEE PAGES 84-85 OF NOTEBOOK "OPTICAL LATTICES -2"
%\vspace{-5mm}
%\begin{figure}[h]
%%\begin{figure}[H]
%%\centering
%\includegraphics{fig3dlattice_no_probe.pdf}
%%\vspace{-6mm}
%\caption{\footnotesize{3D tetrahedral lin$\perp$lin lattice configuration. Beams $\vec{k}_{1}$, $\vec{k}_{2}$ ($\vec{k}_{3}$, $\vec{k}_{4}$) propagate in the \emph{x-z} (\emph{y-z}) plane with polarization along the $\hat{y}$ ($\hat{x}$) direction.}}
%\label{fig:3dlat}
%%\vspace{-2mm}
%\end{figure} 
\noindent where $k_x = k_L \,\mbox{sin}\,\theta_x$, $k_y = k_L\, \mbox{sin}\,\theta_y$, and $k_{z} = k_L (\mbox{cos}\theta_x + \mbox{cos}\theta_y)$.
%, and the 3D lattice well-depth $U_0 = -16\hbar{\Delta'_0}/3$~\cite{grynreview}.  
%where $\Delta' = 8\Delta_0'$
%Eqns.~\ref{eq:s1}-\ref{eq:s4} are consistent with relations defined in Ref.~\cite{grynreview} in terms of the light-shift: For the 1D lattice, the well-depth $U_0 = \frac{4}{3}\hbar\Delta_0'$, where $\Delta_0' = \Delta s_0/2$ is the light-shift per beam. 
$\Omega_{X, Y, Z}$ are the $x, y, z$-components of the vibrational frequency $\Omega_V$. Note that the expression for $\Omega_Z$ above is modified so as to be a better approximation for the case of a $F_g \geq 3 \rightarrow F_e = F_g + 1$ atom~\cite{grynreview}.
%, as is the case in our experiments with $^{85}$Rb atoms at the $F_g = 3 \rightarrow F_e = 4$ transition~\cite{grynreview}: 
${\Delta}' = 8{\Delta_0}'$ is the light-shift at a point of circular polarization for a closed transition having a Clebsch - Gordan coefficient equal to 1. For the $F_g = 3 \rightarrow F_e = 4$ transition in $^{85}$Rb, $I_{sat} = 1.64$ mW/cm$^2$ \textcolor{black}{for $\sigma$-light.}

In the experiments, a weak  $y$-polarized probe beam  of the form~\cite{grynreview} 
\begin{equation}
{\vec{E}}_p (z, t) = e^{i\phi} \, \hat{e}_y \, {\cal E}_p \, e^{i(k_pz - \omega_pt)} + c.c.
\end{equation}
\noindent propagating in the $+ z$-direction (${\cal E}_p << {\cal E}_0$) is made incident on the 3D lattice, as depicted in Fig. 1(a). This probe creates a lattice perturbation that propagates in the $\pm x$-directions, as discussed in the paper.
%From Eqn.~\ref{eq:s6} the spatially averaged optical pumping rate $\overline{\Gamma_\pm}$ from $|1\rangle$ to $|2\rangle$ or $|2\rangle$ to $|1\rangle$ is given by
%\renewcommand{\theequation}{S\arabic{equation}}
%\begin{equation}
%\overline{\Gamma_{\pm}} = \frac{4{\Gamma_S}}{9} \, (\overline{ {{\mbox{cos}}^2} {K_x}x + {{\mbox{cos}}^2}{K_y}y })
%\label{eq:gamma_av}
%\end{equation}
%where the $\mbox{cos}\,{2K_+ z}$ term averages to zero. 
%In our geometry, \newline 
%$\overline{ {{\mbox{cos}}^2} {K_x}x + {{\mbox{cos}}^2}{K_y}y } = 1/2 + 1 = 3/2$~\cite{grynpra2003}.

%\vspace{8mm}
%\newpage

\section{{
%$F_{g}=1/2 \rightarrow F_{e}=3/2$ atom in 
%Probe-modulated 1D dissipative optical lattice: 
C: Visualization of 
%Brillouin propagation of 
the dominant propagating mode [1,1] in Fig. 2}}
%: \newline how probe-modulation yields unidirectional motion}
\label{sec:visual}

As discussed in the paper, a new theory~\cite{cubero22} for Brillouin propagation in a 1D section of a 3D lattice decomposes the atomic current into the contributions from the different atomic density wave modes excited by the weak probe. For the dominant [1,1] mode the atomic density wave propagates at the same velocity as the propagating potential perturbation.
%As mentioned in the main paper, the key features behind how efficient directed propagation can arise from random fluctuations may be understood from a simple model of a fictitious $F_g = 1/2 \rightarrow F_e = 3/2$ atom confined in a 1D lin$\perp$lin optical lattice, which is illuminated by a weak probe beam. 

\begin{figure}[t]
%\vspace{-3mm}
%\centering
\includegraphics[width=7.4cm]{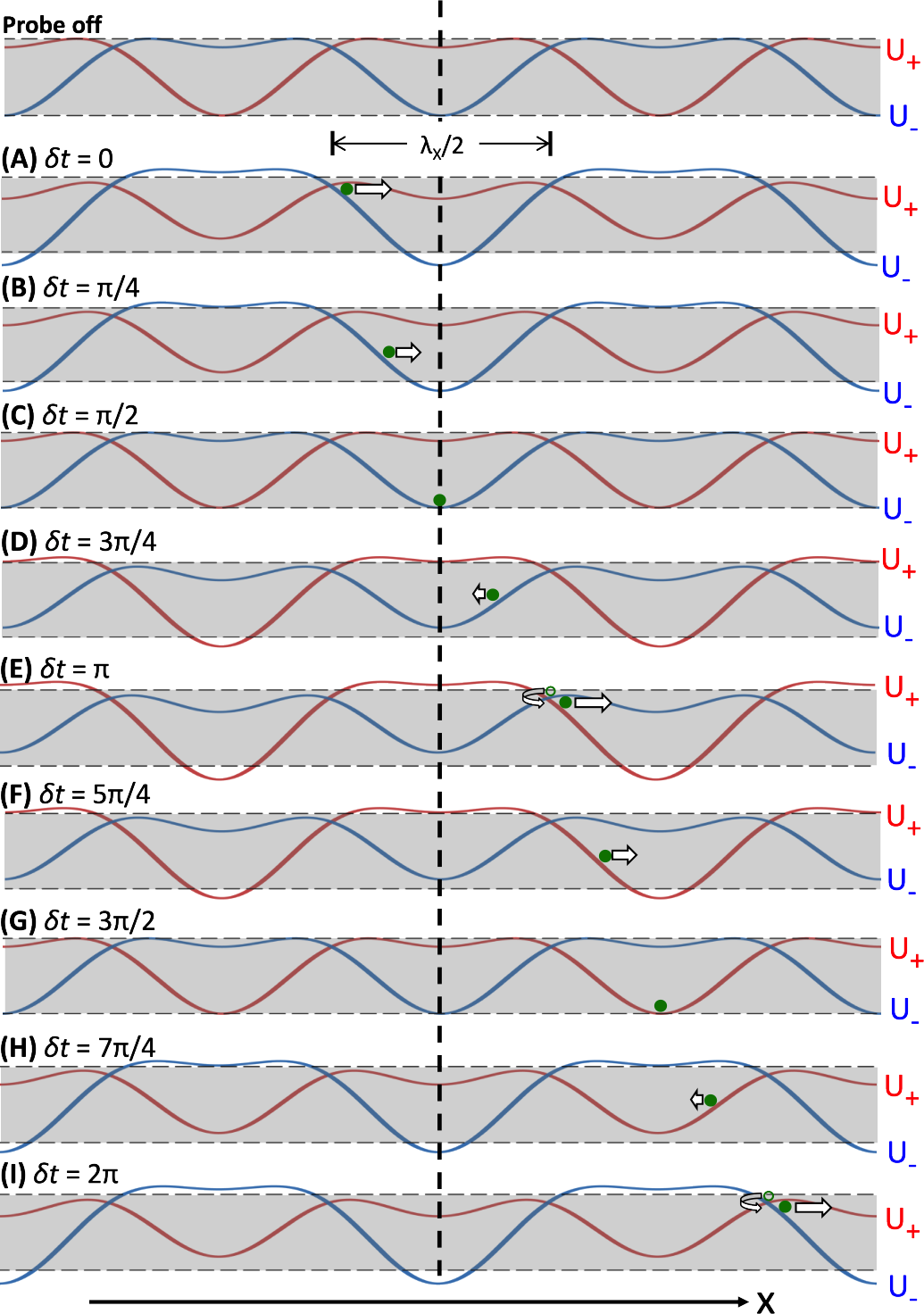}
%\includegraphics{figUpdated_Modulation_2022rev.pdf}
%\vspace{-2mm}
\caption{{{\hspace{0mm} 
Probe-modulated lattice potentials $U_\pm$, obtained by adding ${U_\pm (x)}$ and $U_{\pm}^{p} (x,t)$ in equations (1) and (4) of the main paper. {\bf{(Top)}} Probe OFF: The well-depth of the unperturbed lattice bipotential is demarcated in all the plots by the dashed horizontal lines (shaded gray). The dashed vertical line indicates that the $x$-location for the wells stays fixed. {\bf{(A - I)}} Probe ON: Snapshots of $U_+$ and $U_-$ as they are modulated out-of-phase with each other.
% - when $U_+$ grows deep, $U_(-)$ becomes shallow, and vice versa. 
%they shift side-to-side (shown for one of the $U_+$ wells by dashed vertical arrows), and vary in depth. 
For the [1,1] mode, the time taken by an atom to complete one half-oscillation in the well equals the time taken to complete one half-modulation.
%i.e., $\delta = \Omega_X$. 
For an atom starting at the top left of the $U_-$ well at $\delta t = 0$ as shown, when the well is deepest, and moving with average velocity $+ v_S = \Omega_X/k_x$, the atom experiences, on the average, a deeper well while rolling downhill, from (A) to (C), and a shallower well while rolling uphill, from (C) to (E).   
The white arrows depict the instantaneous force sensed by the atom. The net force forward during the half-cycle (A) to (E) balances the frictional force from Sisyphus cooling, enabling the atom to reach the top on the other side of the well, triggering an optical transition to the adjacent $U_+$ well which is at its deepest at that precise instant. Thus the motion repeats and the atom continues to propagate along $+x$ over many wells. The same directed propagation along $-x$ occurs for an atom starting at the top right of a $U_-$ well at $\delta t = 0$ and moving with average velocity $- v_S$, leading to the creation of a bidirectional ratchet.
%% and accumulate on the right side of the wells. 
%Second, as atoms approach a crossing point,
%%as the atom approaches the crossing point, 
%the tendency to be optically pumped to the adjacent well ($|\Delta \rho_{+} - \Delta \rho_{-}|$) \emph{maximizes} as indicated by the gray solid arrow in (E, I). 
%%Once transferred, the atom continues its motion with the same average velocity $v_B$ but now in the $U_-$ well, and is therefore similarly pumped at $\delta t = 2\pi$ back to the original starting point on $U_+$ (plot I). 
%This process repeats for the $v_B$-velocity class, yielding directed propagation along $+z$ if $\omega_{mod} = \Omega_V$ (see text).
%%This is Brillouin propagation. 
%The number of participating atoms resonantly increases when the optical pumping rate $\Gamma_P$ is synchronized with $\omega_{mod}$ and $\Omega_V$ enabling efficient transfer at (E) and (I). This is stochastic resonance. 
}}}
%\vspace{-5.5mm}
\label{fig:expl}
\end{figure}

A depiction, as in Fig.~\ref{fig:expl}, of the [1,1] mode offers instructive insights into how directed atomic propagation may arise within a sample of randomly diffusing atoms.
The topmost plot shows the unperturbed bipotential $U_\pm (x)$ from (1) when the probe is off. The gray region marks the sizes of the unperturbed well-depths, and the dashed vertical line marks the $x$-location of a particular well. The probe-modulated ground state potentials, denoted in the figure by $U_\pm$, are simply obtained by adding the expressions for $U_\pm (x)$ in (1) and $U_\pm^p (x,t)$ in (4).
 %We use $|{\cal E}_{p\perp}/{\cal E}_0|^2 = 25\%$ to exaggerate probe effects for clarity.
We use $\epsilon_p = {\cal E}_p/{4\cal E}_0 = 0.1$, same as in Fig.~\ref{fig:david}. This value of $\epsilon_p$ is more than twice the maximum value attained in our experiments, and is used to exaggerate probe effects in Fig.~\ref{fig:expl} for visual clarity. 
As indicated in (\ref{eq:cos:part}), the probe contributes two equal and opposite perturbations traveling in the $\pm x$-directions. The result is that the $x$-location of the wells stay fixed, but the depths of adjacent wells oscillate out-of-phase with each other. This is illustrated in Figs. (A) - (I), where successive snapshots of the probe-modulated potentials $U_\pm$ are plotted at $\delta t = 0, \pi/4, \pi/2$, and so on. Note that during the first half of the probe modulation cycle $\delta t = 0 \rightarrow \pi$ (Figs. (A) - (E)), $U_-$ becomes shallower while $U_+$ grows deeper, and the roles are reversed during the second half $\delta t = \pi \rightarrow 2\pi$ (Figs. (E) - (I)).
  
The vibrational frequency $\Omega_X$ coincides with the modulation frequency $\delta$ for the [1,1] mode. Atoms residing near the bottoms of wells are excited by this resonance, causing them to approach regions near the top of the wells where the probability for a transition to the adjacent well by optical pumping is larger. Consider an atom starting at the top of a $U_-$ well, at $\delta t = 0$ when the well is deepest, on the left side as shown and moving with average velocity $+ v_S = \Omega_X/k_x$. Because $\delta = \Omega_X$, the atom completes one half-oscillation in the well in the same time $\delta t = \pi$ that the modulation completes one half-cycle. A key point is that the well is on the average deeper while the atom is rolling downhill, and shallower while rolling uphill. In Fig.~\ref{fig:expl}, the white arrows depict the size and direction of the force at each instant on the atom (not the direction of motion of the atom; this atom is always moving to the right). This asymmetry results in a net force forward on the atom during its half-oscillation, which suppresses the effect of the ever-present frictional force due to Sisyphus cooling. The atom is thus able to reach the top of the well on the other side, triggering a transition to the adjacent well $U_+$, which at $\delta t = \pi$, happens to be at its deepest, and the cycle repeats. The atom can undergo directed propagation in the $+ x$-direction over many successive wells~\cite{grynpra2003,grynprl1996,grynprl2002,grynbergepl}. Of course, an atom starting at the top right of a $U_-$ well, at $\delta t = 0$ when the well is deepest, and moving with average velocity $- v_S$, experiences exactly the same type of directed propagation over many wells in the $-x$-direction. Thus, we have a bidirectional ratchet.

\section{{D: $\Omega_B$ in Fig.~\ref{fig:spectraz} coincides with $\pm \Omega_X$ but is evidence for directed propagation, not intrawell oscillation  
}}
\label{sec:omb_omx}
%Figs. 2 and 3 in the main manuscript, and the accompanying text, describe in the context of a 1D lattice model how probe-induced modifications of the light-shifted ground state lattice potentials on one hand and the optical pumping rates on the other combine remarkably to create unidirectional propagation. 
%In the paper, experimental evidence for directed propagation in the form of a bidirectional ratchet along $\pm x$ in the 3D lattice of Fig.~\ref{fig:fig1}(a), manifests in the probe transmission spectrum shown in Fig.~\ref{fig:spectraz} as the resonances denoted by $\Omega_B$. 
%These resonances appear in addition to the expected features $\Omega_Z$ which denote the $z$-component of the vibrational frequency for the 3D lattice. 

The $z$-propagating probe measures $\Omega_Z$, but cannot measure $\Omega_X$ or $\Omega_Y$. It is important to make this point because $\Omega_B$ happens to coincide with the calculated value for $\Omega_X$. %(since the directed transport happens along the $x$-direction).   
%\section{\large{The $Z$-probe ``shakes" atoms along $z$, hence measures $\Omega_Z$ but not $\Omega_{X}$ or $\Omega_{Y}$}}
%\label{sec:shake}
%\textcolor{black}{Brief proof: z-probe operator is linear in z, but quadratic in x, therefore can \emph{only} detect $\Omega_Z$ because adjacent vibrational levels in a well have opposite parity, and cannot measure $\Omega_X$} 
%The Z-probe is taken to be of the form $\vec{\cal E}_p (z, t) = \frac{1}{2} \hat{e}_p{\cal E}_p e^{i\phi} e^{i(k_p z - \omega_p t)} + c.c.$, where the polarization $\hat{e}_p$ is chosen to be either $\hat{e}_x$ or $\hat{e}_y$, and $\phi$ is chosen as above~\cite{origin}.
Fig. 3(a) in the paper shows a Z-probe transmission spectrum for a specific lattice well-depth. In order for a peak or a dip at a vibrational frequency component to arise in the probe transmission spectrum there must be a probe-induced Raman transition between adjacent intrawell quantized vibrational levels (say $\phi_n$ and $\phi_{n+1}$, which have opposite parity). Since the probe ${\vec {E}}_p$
% $= {\hat{e}_y}{\cal E}_{p} \mbox{sin} (kz - \omega_{p} t)$ 
is weak, the dominant probe term in the total electric field is the lattice-probe interference term which goes as $\vec{E}_0 \cdot {\vec{E}^{*}}_p$ (see Appendix B). The Raman transition is proportional to the overlap integral $\int d\vec{r} \, \phi^{*}_{n+1} \vec{E}_0 \cdot {\vec{E}^{*}}_p \, \phi_{n}$. The product of the wavefunctions is always odd, therefore this integral is nonzero only if the interference term is odd.
%For the case of the $x$-polarized Z-probe, the interference term goes as $\mbox{cos}(ky \, \mbox{sin}\theta_y) \, \mbox{exp} [-ikz(1+\mbox{cos}\theta_y)]$, which for small values of $y$ and $z$ is linear in $z$, but quadratic in $y$. Similarly, 
For the $y$-polarized Z-probe, the interference term goes as $\mbox{cos}(k\,x \, \mbox{sin}\,\theta_x) \, \mbox{exp} [-ik\,z(1-\mbox{cos}\,\theta_x)]$, which for small values of $x$ and $z$ is linear in $z$, but quadratic in $x$ (we have taken $k_L = k_p = k$). Thus, the integrand of the overlap integral is even in $z$ allowing for the detection of Raman transitions at $\Omega_Z$ in Fig. 3(a). However, the observed peak-dip features at $\Omega_X$ in Fig. 3(a) cannot arise from Raman transitions between adjacent vibrational levels because the integrand is an odd function in $x$. We explain in the paper that these spectral features arise from directional propagating modes along $\pm x$.
%, and the observed resonant features $\Omega_B$ coincide with $\Omega_X$ for very different reasons.

\section{{E: Curve-fitting method to obtain $\cal A$}}
\label{sec:curvefit}
\begin{figure}[h]
%\centering
\includegraphics[width=6.2cm]{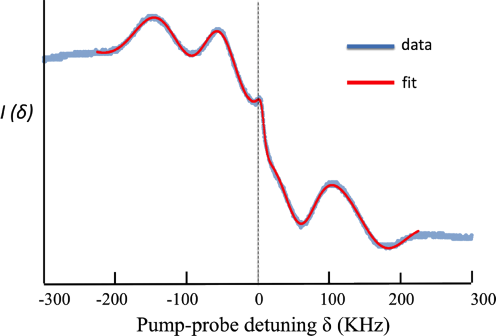}
% \vspace{0mm}
\caption{\footnotesize{The fit function $I(\delta)$ in (\ref{eq:fit}) is used to model the probe transmission spectrum.}}
\label{fig:fit}
%\vspace{-3cm}
\end{figure}

\noindent The probe transmission spectrum in Fig.~\ref{fig:spectraz}(a) is fit by a function $I (\delta)$, as shown in Fig.~\ref{fig:fit}.

The fit function $I(\delta)$, given below, contains four Gaussian functions, two for the spectral peak and dip at the vibrational frequency $\pm \Omega_Z$ and two for the Brillouin peak/dip at $\pm \Omega_B$. $I(\delta)$ also contains Lorentzian and dispersive curves to help fit the central feature of the spectrum, referred to as the ``Rayleigh" feature in the literature~\cite{grynreview}, believed to arise from atomic velocity damping due to Sisyphus cooling~\cite{grynprl2003}. Finally, a linear and constant term are included to further improve the fitting.
\setcounter{equation}{0}
\renewcommand{\theequation}{E\thesection\arabic{equation}}
\begin{eqnarray}
I(\delta) & = & A_{Z1}{\mbox e}^{-\frac{(\delta - \Omega_{Z1})^2}{2{\sigma_{Z1}}^2}} + A_{Z2}{\mbox e}^{-\frac{(\delta - \Omega_{Z2})^2}{2{\sigma_{Z2}}^2}} 
\nonumber \\
& & + A_{B1}{\mbox e}^{-\frac{(\delta - \Omega_{B1})^2}{2{\sigma_{B1}}^2}} + A_{B2}{\mbox e}^{-\frac{(\delta - \Omega_{B2})^2}{2{\sigma_{B1}}^2}} \nonumber \\
& & + a_1 + a_2 \delta + \frac{a_3}{(\delta + x_0)^2 + {\gamma}^2} \nonumber\\
& & + \frac{a_4 (\delta + x_0)}{(\delta + x_0)^2 + {\gamma}^2}
\label{eq:fit}
\end{eqnarray} 
%\end{appendices}
% Create the reference section using BibTeX:
%\bibliography{basename of .bib file}

%%\vspace{-6mm}
%\section{Acknowledgements}
%\begin{acknowledgements}
%%%\vspace{-4mm}
%This work is supported by the Army Research Office under award/contract number W911NF2110120.  {\dcor DC acknowledges financial support from the Ministerio de Ciencia e Innovaci\'on of Spain of Spain, Grant No. PID2019-105316GB-I00.} We thank the Instrumentation Laboratory at Miami University for electronics and LabView support. 
%We thank A. Dharmasiri and A. Rapp for assistance during the initial setup. 
%We are grateful to A. Reinhard and N. Peters for providing invaluable feedback on the manuscript.
%\end{acknowledgements}
\end{document}